\documentclass[aoas,preprint]{imsart}

\RequirePackage[OT1]{fontenc}
\RequirePackage{amsthm,amsmath,gensymb}
\RequirePackage[authoryear]{natbib}
\RequirePackage[colorlinks,citecolor=blue,urlcolor=blue]{hyperref}

\usepackage[sc]{mathpazo}
\usepackage[T1]{fontenc}
\usepackage{geometry}
\usepackage{appendix}

\usepackage{comment}

\usepackage{url}
\usepackage{multirow}

\usepackage{amsmath}
\usepackage{mathrsfs}
\usepackage{mathtools}
\usepackage{eucal}
\usepackage{amsfonts}
\usepackage{amssymb}
\usepackage{natbib}
\usepackage{hyperref}
\usepackage{hypernat}
\usepackage{times}
\usepackage{epsfig}
\usepackage{booktabs}
\usepackage{rotating}
\usepackage{cancel}
\usepackage{subfigure}
\usepackage{arydshln}

\arxiv{arXiv:0000.0000}

\startlocaldefs
\numberwithin{equation}{section}
\theoremstyle{plain}

\endlocaldefs

\newcommand{\bmat}{\begin{pmatrix}}
\newcommand{\emat}{\end{pmatrix}}

\newcommand{\vb}[1]{{\verb|#1|}}

\newcommand{\bi}{\begin{itemize}}
\newcommand{\ei}{\end{itemize}}
\newcommand{\benum}{\begin{enumerate}}
\newcommand{\eenum}{\end{enumerate}}
\newcommand{\bv}{\begin{verbatim}}
\newcommand{\ev}{\end{verbatim}}
\newcommand{\bm}[1]{\mbox{\boldmath $#1$}}

\newcommand{\Rprog}{\textsf{R}} 
\newcommand{\R}{\mathbb{R}} 
\newcommand{\non}{\nonumber}

\newcommand{\sst}{\textrm{sst}}

\newcommand{\E}{\textrm{E}} 
\newcommand{\cov}{\textrm{cov}} 
\newcommand{\pP}{\textrm{P}} 
\newcommand{\pN}{\textrm{N}} 
\newcommand{\pPo}{\textrm{Po}} 
\newcommand{\pLN}{\textrm{logNormal}} 

\newcommand{\li}{\lambda} 
\newcommand{\wt}[1]{\widetilde{#1}}
\newcommand{\wh}[1]{\widehat{#1}}
\newcommand{\mv}[1]{\boldsymbol{#1}}
\newcommand{\ol}[1]{\overline{#1}} 
\newcommand{\CC}{\mathcal{C}} 

\newcommand{\mmd}{{\mathrm{d}}} 
\newcommand{\md}{{\,\mmd}} 

\theoremstyle{remark}
\newtheorem*{remark}{Remark}

\renewcommand\appendixname{Supplement}

\begin{document}

\begin{frontmatter}
\title{Point process models for spatio-temporal distance sampling data from a large-scale survey of blue whales}
\runtitle{Point process models for spatio-temporal distance sampling data}
\thankstext{T1}{The project is funded by the Engineering and Physical Sciences Research Council(EPSRC) --EP/K041061/1 and EP/K041053/1.}

\begin{aug}
\author{\fnms{Yuan} \snm{Yuan}\thanksref{m1}\ead[label=e1]{yy84@st-andrews.ac.uk}},
\author{\fnms{Fabian E.} \snm{Bachl}\thanksref{m2}\ead[label=e2]{fabian.bachl@ed.ac.uk}},
\author{\fnms{Finn} \snm{Lindgren}\thanksref{m2}\ead[label=e3]{finn.lindgren@ed.ac.uk}},
\author{\fnms{David L.}
\snm{Borchers}\thanksref{m1}\ead[label=e4]{dlb@st-andrews.ac.uk}},
\author{\fnms{Janine B.} \snm{Illian}\thanksref{m1, m3}\ead[label=e5]{jbi@st-andrews.ac.uk}},
\author{\fnms{Stephen T.} \snm{Buckland}\thanksref{m1}\ead[label=e6]{steve@st-andrews.ac.uk}},
\author{\fnms{H{\aa}vard} \snm{Rue}\thanksref{m4}\ead[label=e7]{haavard.rue@kaust.edu.sa}},
\and
\author{\fnms{Tim} \snm{Gerrodette}\thanksref{m5}\ead[label=e8]{tim.gerrodette@noaa.gov}}

\runauthor{Y. Yuan et al.}

\affiliation{University of St Andrews\thanksmark{m1}, University of Edinburgh\thanksmark{m2}, Norwegian University of Science and Technology\thanksmark{m3}, King Abdullah University of Science and Technology\thanksmark{m4}, and Southwest Fisheries Science Center, NOAA National Marine Fisheries Service\thanksmark{m5}}

\address{Address of Yuan Yuan\\
School of Mathematics and Statistics \\
University of St Andrews\\
The Observatory, Buchanan Gardens \\
St Andrews, UK\\
KY16 9LZ\\
\printead{e1}}

\address{Address of Fabian E. Bachl\\
School of Mathematics\\
University of Edinburgh\\
James Clerk Maxwell Building\\
The King's Buildings\\
Peter Guthrie Tait Road\\
Edinburgh, UK\\
EH9 3FD\\
\printead{e2}}

\address{Address of Finn Lindgren:\\
	School of Mathematics\\
	University of Edinburgh\\
	James Clerk Maxwell Building\\
	The King's Buildings\\
	Peter Guthrie Tait Road\\
	Edinburgh, UK\\
	EH9 3FD\\
	\printead{e3}}

\address{Address of David L. Borchers\\
School of Mathematics and Statistics\\
	University of St Andrews\\
	The Observatory, Buchanan Gardens \\
  St Andrews, UK\\
	KY16 9LZ\\
	\printead{e4}}

\address{Address of Janine B. Illian\\
	School of Mathematics and Statistics\\
	University of St Andrews,\\
	The Observatory, Buchanan Gardens \\
	St Andrews, UK\\
	KY16 9LZ\\
	\printead{e5}}

\address{Address of Stephen T. Buckland\\
	School of Mathematics and Statistics\\
	University of St Andrews\\
	The Observatory, Buchanan Gardens \\
	St Andrews, UK\\
	KY16 9LZ\\
	\printead{e6}}

\address{Address of H{\aa}vard Rue\\
	CEMSE Division\\
	King Abdullah University of Science and Technology\\
        Thuval 23955-6900\\
        Saudi Arabia\\
	\printead{e7}}

\address{Address of Tim Gerrodette\\
	NOAA National Marine Fisheries Service\\
	Southwest Fisheries Science Center\\
	8901 La Jolla Shores Drive\\
	La Jolla, California 92037, USA\\
	\printead{e8}}
\end{aug}

\begin{abstract}
Distance sampling is a widely used method for estimating wildlife population abundance.  The fact that conventional distance sampling methods are partly design-based constrains the spatial resolution at which animal density can be estimated using these methods. Estimates are usually obtained at survey stratum level. For an endangered species such as the blue whale, it is desirable to estimate density and abundance at a finer spatial scale than stratum.  Temporal variation in the spatial structure is also important.  We formulate the process generating distance sampling data as a thinned spatial point process and propose model-based inference using a spatial log-Gaussian Cox process. The method adopts a flexible stochastic partial differential equation (SPDE) approach to model spatial structure in density that is not accounted for by explanatory variables, and integrated nested Laplace approximation (INLA) for Bayesian inference. It allows simultaneous fitting of detection and density models and permits prediction of density at an arbitrarily fine scale.  We estimate blue whale density in the Eastern Tropical Pacific Ocean from thirteen shipboard surveys conducted over 22 years. We find that higher blue whale density is associated with colder sea surface temperatures in space, and although there is some positive association between density and mean annual temperature, our estimates are consitent with no trend in density across years. Our analysis also indicates that there is substantial spatially structured variation in density that is not explained by available covariates. 

\end{abstract}


\end{frontmatter}

\section{Introduction}\label{sec:intro}

Distance sampling is a widely-used set of survey methods for
estimating animal density or abundance \citep{Buckland+al:2001,
  Buckland+al:2015book}. Conventional distance sampling methods
(of which line transect and point transect methods are the most
common) use a combination of model-based inference for estimating
detection probability and design-based inference with
Horvitz-Thompson-like estimators \linebreak \citep{Borchers+al:1998} for
estimating density and abundance conditional on the detection
probability estimates. While the design-based nature of the second
stage in this two-stage estimation process
\citep[see][]{Buckland+al:2016} confers robustness on density and
abundance estimates when suitable designs are used, it severely
restricts the spatial resolution at which such estimates can be
obtained. This is because design-based
inference requires adequate sampling units (strips for
line transect surveys and circular plots for point transect surveys) 
in each area for which animal density or abundance is to be estimated. 
The low spatial resolution of estimates from this two-stage approach
limits the utility of estimates obtained from conventional 
distance sampling methods as there is often interest in the 
distribution at high spatial resolution. As a result, there has been 
increasing interest in distance sampling methods that generate continuous 
spatial density surface estimates, and hence allow inference at 
an arbitrarily fine spatial scale.

In this paper, we consider a series of line transect surveys of blue
whales ({\it Balaenoptera musculus}) in the Eastern Tropical Pacific
Ocean \citep[ETP, ][]{Gerridette+Forcada:2005}, in which the focus
of inference is on how density changes continuously in space, with
respect to available explanatory variables, and across
years. The surveys were designed for dolphins, not blue whales, so there are relatively few blue whale sightings. A continuous spatial model has the potential to borrow strength from data outside the lightly-sampled strata to improve overall inference.

One can obtain a continuous density model by using a spatial model of density in the second stage, rather than basing inference on the design in this stage. This is usually done by transforming the data to counts: discretizing the sampled strips into smaller spatial units in the case of line transects and specifying a model for the counts within each unit, using estimated detection probability as an offset to correct the counts for detectability. \citet{Hedley+al:2004} and \citet{Hedley+Buckland:2004} pioneered this approach and \citet{Niemi+Fernandez:2010} developed a similar approach (but ignoring detection uncertainty). The \Rprog-package \texttt{dsm} \citep{Rpack:dsm} implements the approach of \citet{Hedley+al:2004} and \citet{Hedley+Buckland:2004} using generalized additive models
\citep[GAMs,][]{Wood:2006} to estimate a density surface from the count data. Either frequentist or Bayesian
approaches can be used for the second stage \citep{Oedekoven+al:2013,Oedekoven+al:2015}, and bootstrapping is often used to propagate the uncertainty of detectability estimated from the first stage. \citet{Williams+al:2011} use a more direct approach to incorporate uncertainty of detectability: a random effect term is added in the second stage to characterize the uncertainty in the estimation of the detection function from the first stage.

One can also estimate the parameters of the detection function and the count model simultaneously \citep{Royle+Dorazio:2008, Royle+al:2004, Johnson+al:2010, Moore+Barlow:2011, Schmidt+al:2012, Conn+al:2012,   Oedekoven+al:2014, Pardo+al:2015}. This is known as a full-likelihood approach, as it involves specifying a likelihood that incorporates both a detection function model and a spatial density
model, allowing simultaneous estimation of both models.

Whether inference is in two stages or one, models that discretize
searched strips or lines involve an element of subjectivity in
choosing the size of the discrete units and a loss of spatial
information because each discrete unit can have only one value of any
spatial covariate attached to it, even though it might span an area
incorporating a range of covariate values. In this paper, we develop a
method that does not suffer from these problems, using a
point process model.

Point process theory provides a flexible modeling framework for
incorporating the underlying spatial or spatio-temporal stochastic
processes and does not require discretization of spatial sampling
units. Point process models have been used with ecological data to
estimate smooth spatial density surfaces and are an obvious choice for
the spatial model component of a full likelihood line transect model,
although to date they have mainly been used in ecological applications
with fully mapped point patterns \citep{Wiegand+Moloney:2014}: \citet{Stoyan:1982} formulated
line transect data as observations of stationary point
processes; \citet{Hedley+al:2004} considered point process models
for point transect surveys, and \citet{Hogmander:1991, Hogmander:1995}
constructed a marked point process model for line transect data with
detection probability of an animal treated as a mark, but they used
a detection model (in which each animal has a detection circle with
variable radius) that was shown by \citet{Hayes+Buckland:1983} to be
unrealistic and often resulting in biased inference.

Here we develop a full likelihood point process model for line
transect data, in which the detection process thins the underlying
point process, and in which the detection model and the point process
model are estimated simultaneously. In the context of the blue whale survey, an unknown point process governs the number and locations of the whales in space, and points are thinned (whales missed) with a probability that depends in an unknown way on distance from the known locations of lines. Such an approach is not new for
modeling distance sampling data. The \Rprog-package \texttt{DSpat}
\citep{Rpack:DSpat, Johnson+al:2010} uses a thinned point process
model for line transect survey data.
However, their method assumes the absence of residual spatial structure
on the intensity level (whale density in our case), which is usually not the case in practice, and may result
in biased estimates. We relax the independence assumption by using the
stochastic partial differential equation approach \citep[SPDE,
][]{Lindgren+al:2011} to incorporate a spatial or spatio-temporal
random field for the underlying stochastic process of autocorrelated
spatial or spatio-temporal random effects. For point process data in
general, the SPDE approach avoids the need to aggregate observations
\citep{Simpson+al:2015}, and it provides a flexible modeling framework
for spatio-temporal random fields. We build our models in a Bayesian
framework, which gives us a tool for fitting complicated models, and
the advantage of being able to use integrated nested Laplace approximation
\citep[INLA,][]{Rue+al:2009} for inference. INLA is a computationally efficient
method for Bayesian inference using numerical approximations instead
of a sampling-based method such as Markov chain Monte Carlo
algorithms. In addition, our modeling framework accommodates the models
of the sort used by \citet{Johnson+al:2010} as a special case, as well as the second
stage of the two-stage approach of \citet{Miller+al:2013}.

After describing the blue whale survey data in Section~\ref{sec:surveydata}, we describe our model and computational methods in Sections~\ref{sec:modelsoverall} and Section~\ref{sec:computation}. We then analyze the surey data in Section~\ref{sec:casestudy}, investigating the underlying spatial stochastic process of blue whale density in this area, and how the blue whales respond to sea surface temperature in space and time. Finally, in Section~\ref{sec:discuss}, we discuss the results of the analysis, the utility of our modeling approach and extensions for more complicated scenarios.

\section{The blue whale survey data\label{sec:surveydata}}

Line-transect cetacean surveys were carried out in the Eastern Tropical Pacific Ocean (ETP) between 1986 and 2007. Fig~\ref{fig:effortsight} shows the survey region and transect lines over this whole period, together with blue whale sightings. The survey area is 21.353 million square kilometres and is large enough that the curvature of the earth needs to be taken into account in the analysis. A total of 182 blue whale groups were sighted over all years, with a mean group size of 1.8 (standard deviation 2.1). In 1986-1990, 1998-2000, 2003 and 2006, the entire ETP area of was sampled. These complete surveys required two oceanographic research vessels (3 in 1998) for 120 sea days each. Transect search effort was stratified by area \citep{Gerridette+Forcada:2005}, and in 1992, 1993 and 2007, only part of the ETP area was sampled.  These spatial differences in intensity of sampling need to be accounted for in modelling (see Section~\ref{sec:numericalintegration} for more detail). Data collection followed standardized line-transect protocols \citep{Kinzey+al:2000}. Briefly, in workable conditions, a visual search for cetaceans was conducted by a team of three observers on the flying bridge of each vessel during all daylight hours as the ship moved along the transect at a speed of 10 knots.  Pedestal-mounted 25X binoculars were fitted with azimuth rings and reticles for angle and distance measurements. If a blue whale sighting was less than $10$ km from the transect, the team went off-effort  and directed the ship to leave the transect to approach the sighted animal(s). The observers identified the sighting to species or subspecies (if possible) and made group-size estimates. 

The inference problem we address is how to model the density of blue whale groups across this survey region in a way that takes account of (i) the variable survey effort (transect lines) in space, (ii) the unknown probability of detecting a group from a line, with detection probability decreasing with distance from line, (iii) the dependence of density on explanatory variables (sea surface temperature in particular), (iv) how density changes over years, and (v) spatial fluctuation in blue whale density that cannot be explained by any available explanatory variables.

We describe the statistical models and tools that we use to address this inference problem next, and then use these to address the blue whale inference problem.

\section{The models}\label{sec:modelsoverall}

\begin{figure}[h!]
\centering
\begin{tabular}{cc}
\includegraphics[width=0.4\textwidth]{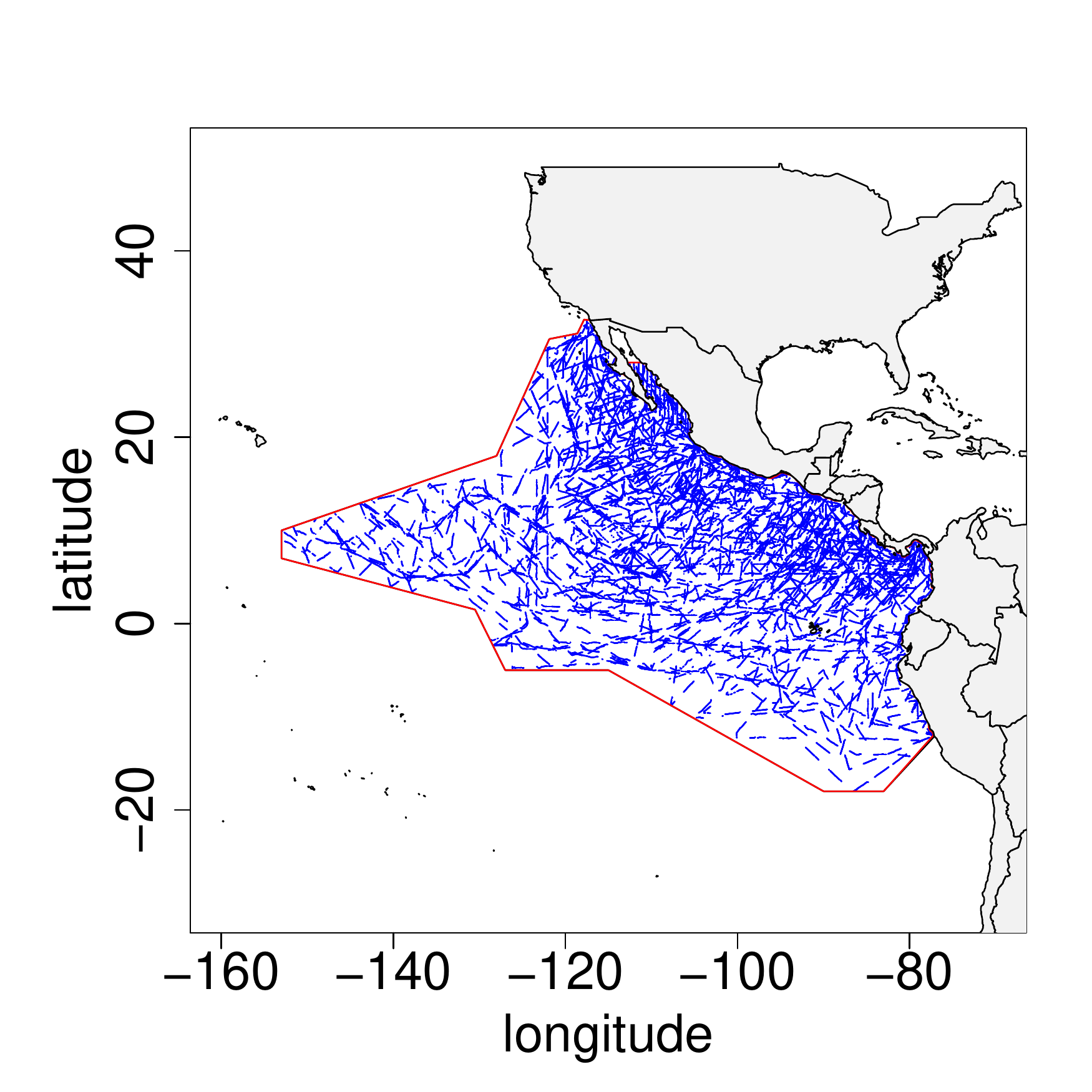}&
\includegraphics[width=0.4\textwidth]{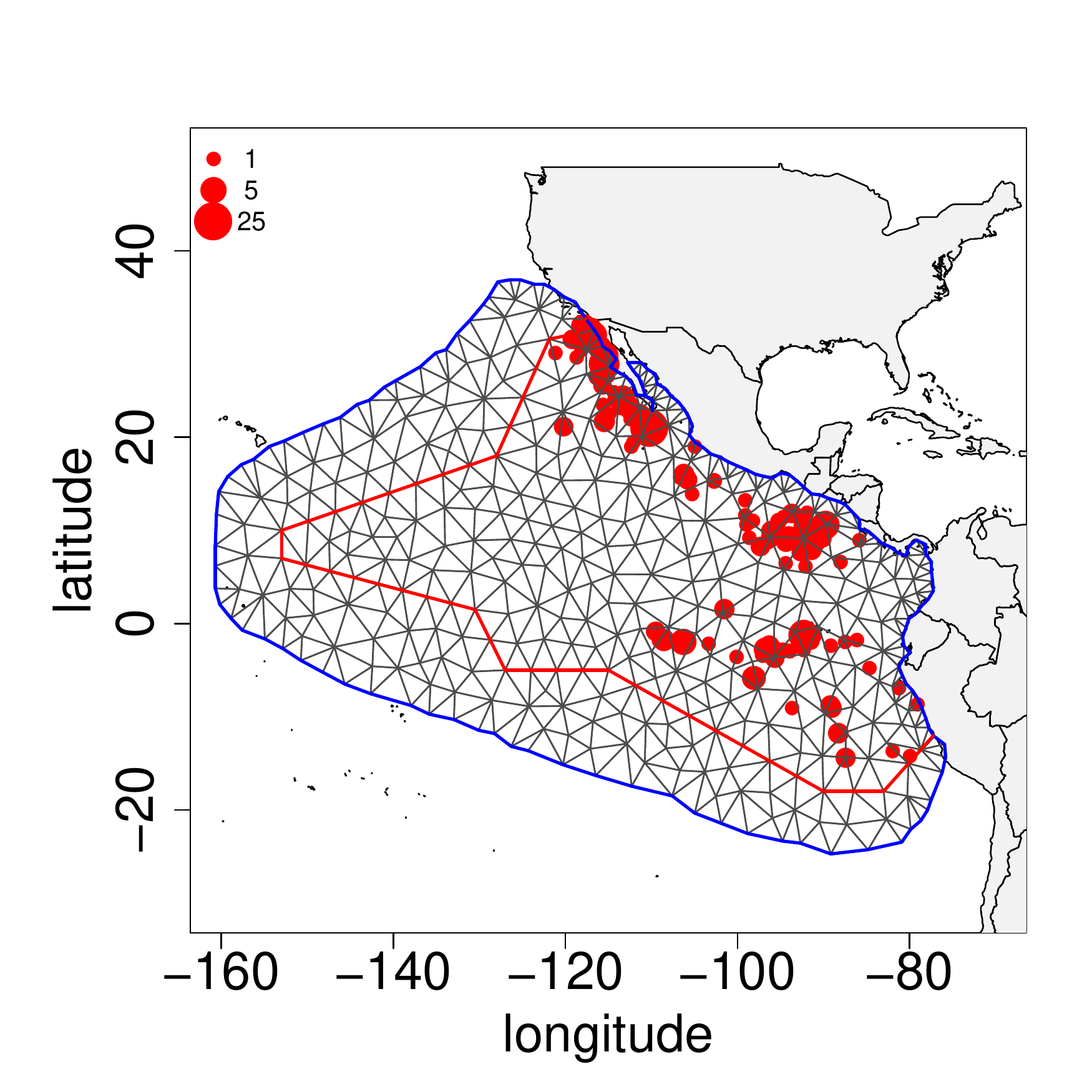}
\end{tabular}
\caption{Plot of the ETP data: the left panel shows the transect lines surveyed from 1986 to 2007, and the right panel displays the sightings of blue whale groups (red dots) on top of the mesh used in our analysis: the radius of each dot is proportional to the logarithm of the group size plus 1. The red line is the ETP survey region boundary.  }
\label{fig:effortsight}
\end{figure}

Spatial point processes model the locations of objects in space
\citep{Stoyan+Grabarnik:1991, Lieshout:2000, Diggle:2003, Moller+Waagepetersen:2004, illianBook}. Before incorporating distance sampling, we consider spatial point patterns formed by objects, represented as collections of locations,
$\bm{Y}\equiv\{\bm{s}_i,\, i = 1, \dots, n\}$. The point set $\bm{Y}$
is considered as a realisation from a random point process on a
bounded domain $\Omega$, where usually $\Omega\subset\R^2$.
Since the ETP survey domain is large enough for the curvature of the
Earth to matter (see Fig~\ref{fig:effortsight}), we treat $\Omega$ as a subdomain of a sphere, $\Omega\subset\mathbb{S}^2$.

\subsection{Spatial hierarchical Poisson point process models}
\label{sec:hierarchical}
For any subset $A\subseteq\Omega$, the number of objects in $A$ is
denoted $N_{\bm{Y}}(A)$.  For an inhomogeneous point process, we
define an intensity function $\Lambda(\bm{s})$ as
\begin{align*}
  \Lambda(\bm{s}) &= \lim_{\epsilon\rightarrow 0}
  \frac{
    \E\{N_{\bm{Y}}[\mathcal{B}_\epsilon(\bm{s},t)]\}
  }{|\mathcal{B}_\epsilon(\bm{s},t)|},\quad \bm{s}\in\Omega,
\end{align*}
where $\mathcal{B}_\epsilon(\bm{s})$ is a ball of radius $\epsilon$ centered at $\bm{s}$. For all non-overlapping subsets $A_1,\dots,A_m\subset\Omega$, an inhomogeneous Poisson point process has the following two conditions,
\begin{align*}
  &N_{\bm{Y}}(A_k) \sim \pPo\left[\int_{A_k} \Lambda(\bm{s}) \md\bm{s} \right],\quad \text{$k=1,\dots,m$, and}
  \\
  &\text{$N_{\bm{Y}}(A_1),\dots,N_{\bm{Y}}(A_m)$ are mutually independent.}
\end{align*}
Finally, we let $\Lambda(\bm{s})$ be a random process, and define the
point pattern model conditionally on
$\Lambda(\bm{s})$.  The conditional likelihood for the
entire point pattern $\bm{Y}$, relative to a homogeneous Poisson process with intensity $1$, is given by
\begin{align}
\label{eq:lgcplikelihood}
\pi(\mathbf{Y} | \Lambda) &=  \exp\left( |\Omega| - \int_{\Omega} \Lambda(\bm{s}) \md\bm{s}\right) {\prod_{i=1}^{N_{\bm{Y}}(\Omega)}\Lambda(\bm{s}_i)},
\end{align}
where $\bm{s}_i$ is the location of the $i$th observation. If $\log\Lambda(\bm{s})$ is modeled by a latent Gaussian linear model, the resulting hierarchical model is a doubly-stochastic
log-Gaussian Cox process \citep{Moller+Waagepetersen:2004}.

\subsection{Point process models in the context of distance sampling}
\label{sec:PPmodelsDist}
For wildlife surveys, only a proportion of the population in the
domain of interest is observed, due to partial sampling of the domain, and failure to detect all animals in the sampled regions. Distance sampling provides a method to account for imperfect detection. In line transect surveys, an observer traces a path through space,
searching a strip centered on the path.  The probability of detecting an object typically decreases with distance from the observer.  From a modeling perspective, this results in a \emph{thinned} spatial point process with the intensity function scaled by the detection probability \citep{Dorazio:2012,Johnson+al:2013,Hefley+Hooten:2016}.

When deriving the appropriate likelihood model for an observed point
pattern, the problem-specific underlying generative structure
influences the potential dependence between point locations both over
space and in time. It is therefore important to note that the thinning in transect surveys is
neither a thinning of a fixed spatial point pattern, nor a thinning of a regular spatio-temporal point process.  Instead, each object is
characterised by a temporally evolving curve in space, describing its movement, and the observations are thinned snapshots of time-slices of the resulting point process of curves.  In addition, the intensity may vary over time, and we write $\lambda(\bm{s};t)$ for the spatial point intensity for the full time-slice point pattern at time $t$, and $\Lambda(\bm{s};t)$ for the intensity of the observationally thinned version. The assumptions about the movements of the observer and the objects affect what approximations are allowed in practical calculations.

Traditionally, the detection probability for an object located at a
given perpendicular distance $z$ from the path of the observer is
modeled by a \emph{detection function} $g(z)$. Assuming that the observer is moving with constant speed $v$ along a straight line, standard Poisson process theory yields the probability of detecting an object located at $\mv{s}_0$ as a function of the perpendicular distance $z(\mv{s}_0)$,
\begin{align*}
  \pP(\text{object at $\bm{s}_0$ detected} \mid \bm{s}_0\in\bm{Y}) &=
  1 - \exp\{- h[z(\bm{s}_0)]/v\}
  =
  g(z(\bm{s}_0),v),
\end{align*}
where $h(\cdot)$ is an aggregated detection \emph{hazard} along the
path, and $g(\cdot,\cdot)$ is the aggregated detection function, with
explicit dependence on $v$.  The standard approach is to model either
the aggregated detection probability $g(z,v)$, or the aggregated
hazard $h(z)$.  Under simple assumptions about the observers,
\citet{Hayes+Buckland:1983} derived the commonly-used \emph{hazard-rate model}, given by $h(z)=-(z/\sigma)^{-b}$, $b,\sigma>0$.  The
half-normal detection function
$g(z)=\exp\left[-z^2/(2\sigma^2)\right]$, $\sigma > 0$ is another widely-used model.  While the hazard-rate model is more flexible than the half-normal detection
model, only the latter results in a log-linear probability model.  For
this reason, the hazard-rate model does not fit directly into the
existing INLA estimation software \citep{Rue+al:2009}, and instead we use a semi-parametric detection model, which we introduce in
Section~\ref{sec:positivegx}, to give us a more flexible model than the half-normal.

\subsection{Line transect point process likelihood}
\label{sec:linetransectlikelihood}

For line transects, assuming that environmental and other
observational conditions that might affect detectability remain
constant along suitably short and straight \emph{transect segments},
we can formulate a tractably simple version of the
likelihood.  The region of space swept by the transect path is assumed
to consist of a sequence of rectangular transect segment strips
$\{\CC_1,\dots,\CC_K\}$, so that $\CC_{k(t)}$ is the transect strip at time $t$.  Writing $\lambda(\mv{s};t)$
for the intensity of potentially observable objects, and introducing
transect-dependent detection functions $g_{k(t)}(\bm{s})$, the intensity for
the thinned observational point process is
$\Lambda(\bm{s};t)=\lambda(\bm{s};t) g_{k(t)}(\bm{s})$.
Under some loose assumptions (see Supplement~\ref{appx:LTassumptions}), the joint conditional likelihood for the observed point pattern is the product of the conditional likelihoods
for each individual transect segment,
\begin{align}
  \label{eq:likelihood}
  \pi(\bm{Y} | \Lambda) &=
  \exp\left( \sum_{k=1}^K |\CC_k| -
\sum_{k=1}^K \int_{\CC_k} \Lambda(\bm{s}; t_{\CC_k})
d\bm{s}\right) {\prod_{i=1}^{N_{\bm{Y}}} \Lambda(\bm{s}_i; \,t_i)},
\end{align}
where $N_{\bm{Y}}=\sum_{k=1}^K N_{\bm{Y}}(\CC_k)$ is the total
number of observed objects, located at $(\bm{s}_i,t_i)$, $i=1,\dots,N_{\bm{Y}}$. We do not specifically address the issue of \emph{marks} here (features or quantities associated with detected groups or animals). Marks that do not affect the detection probability can
be modeled alongside the object intensity $\lambda(\bm{s};t)$,
including possible common fixed effects and dependent random effects
\citep{Illian+al:2012}. However, marks that do affect the detection probability, such as the sizes of groups of animals, require a joint likelihood expression for
the extended dimension point process of object locations and their
marks, which is a topic for further development.

\subsection{A Bayesian hierarchical spatio-temporal point process model}
\label{sec:model}
Following the classical approach for log-Gaussian Cox processes, we
let the logarithm of the intensity $\lambda(\bm{s};t)$ be a Gaussian
process, with linear covariates $\bm{x}(\bm{s}, t)$, and a zero mean
additive Gaussian spatial or spatio-temporal random field
$\xi(\bm{s},t)$ \citep{Moller+al:1998,Moller+Waagepetersen:2004,Moller+Waagepetersen:2007}.  For computational efficiency, we use the INLA method for numerical Bayesian inference with Gaussian
Markov random fields \citep{Rue+al:2009, Illian+al:2012, Simpson+al:2012}, but the
general methodology is not tied to a specific inferential framework.

In the likelihood given by (\ref{eq:likelihood}), the log of the thinned intensity
is given by
\begin{align}
\label{eq:logLambda}
\log[\Lambda(\bm{s}; t)]=\log[\lambda(\bm{s}; t)] + \log [g_{k(t)}(\bm{s})]=\bm{x}(\bm{s}, t)^\top\bm{\beta} + \xi(\bm{s},t)+ \log [g_{k(t)}(\bm{s})],
\end{align}
where we assume Gaussian priors for $\bm{\beta}$, and a Gaussian random
field $\xi$.  If the logarithm of the detection probability model is
linear in its parameters, this results in a joint linear model with
latent Gaussian components.

In general, any link function and spatially coherent linear predictor
could be used for $\Lambda$.  The point process likelihood only
requires $\Lambda$ to be well-defined pointwise, and integrable. In
practice, the numerical integration schemes used for practical
likelihood evaluation (see Section~\ref{sec:computation}) require
piecewise continuity and differentiability.  Covariates affecting
$\lambda(\bm{s};t)$ need to be available throughout the transect
region for parameter inference, and throughout the domain of interest
for spatial prediction. For practical implementation reasons, spatial
covariates are projected onto the same computational function space as
the latent field $\xi$ (see Section~\ref{sec:SPDEcomputation}).
Covariates affecting $g_k(\bm{s})$ need to be available for each
transect segment. Within-segment variation in detectability would
require a more expensive numerical integration scheme in
Section~\ref{sec:SPDEcomputation}, equivalent to splitting segments
until they were sufficiently short for our assumption of constant
detectability within each segment to be fulfilled.  As noted at the
end of Section~\ref{sec:linetransectlikelihood}, marks for individuals
are currently only allowed if they do not affect the detection
probability.

The full model is given by the following hierarchy,
\begin{align*}
  \pi(\bm{Y},\xi,\bm{\beta},\bm{\beta}_g,\bm{\theta}) &=
  \pi(\bm{Y}\mid\xi,\bm{\beta},\bm{\beta}_g,\bm{\theta})
  \pi(\xi\mid\bm{\theta})
  \pi(\bm{\beta}\mid\bm{\theta})
  \pi(\bm{\beta}_g\mid\bm{\theta})
  \pi(\bm{\theta}),
\end{align*}
where $\bm{\beta}_g$ are parameters controlling the detection model,
$\bm{\theta}$ are further model parameters, such as precision
parameters for the latent Gaussian variables.  Each of the prior
densities $\pi(\xi\mid\bm{\theta})$, $\pi(\bm{\beta}\mid\bm{\theta})$,
$\pi(\bm{\beta}_g\mid\bm{\theta})$, and $\pi(\bm{\theta})$ are
controlled by hyperparameters. Note that in some software packages,
including INLA, the parameters $\bm{\theta}$ themselves are referred
to as hyperparameters.

For given prior distributions, the goal is to compute the posterior
densities for the latent variables, optionally with $\bm{\theta}$
integrated out:
\begin{align*}
  \pi(\bm{\theta}\mid\bm{Y})
  &\propto
  \left.
  \frac{
    \pi(\bm{Y},\xi,\bm{\beta},\bm{\beta}_g,\bm{\theta})
  }{
    \pi(\xi,\bm{\beta},\bm{\beta}_g\mid\bm{Y},\bm{\theta})
  }
  \right|_{(\xi,\bm{\beta},\bm{\beta}_g)=(\xi^*,\bm{\beta}^*,\bm{\beta}_g^*)},
  \\
  \pi(\xi,\bm{\beta},\bm{\beta}_g\mid\bm{Y}) &=
  \int
  \pi(\xi,\bm{\beta},\bm{\beta}_g\mid\bm{Y},\bm{\theta})
  \pi(\bm{\theta}\mid\bm{Y})
  \md\bm{\theta},
\end{align*}
where $(\xi^*,\bm{\beta}^*,\bm{\beta}_g^*)$ is an arbitrary latent
variable state vector.  In the INLA method \citep{Rue+al:2009} this is
achieved approximately by replacing
$\pi(\xi,\bm{\beta},\bm{\beta}_g\mid\bm{Y},\bm{\theta})$ with various
Gaussian or near-Gaussian approximations, and integrating numerically
over $\bm{\theta}$.

\subsection{Stochastic PDE models}
\label{sec:basicSPDE}
The general model construction requires no particular assumptions on
how the spatial or spatio-temporal random field $\xi(\bm{s},t)$ is
modeled or treated computationally.  The only requirement is that the
model can be written as a latent Gaussian random field in such a way
that the model likelihood can be evaluated numerically.  In the
context of INLA, that means that we need to construct a Gaussian
Markov random field representation of the continuous space process.
The traditional approach is to discretize space into a lattice and
count the number of sighted points in each lattice cell, but here we
take an alternative approach that allows us to use the true sighting
locations, and to let $\lambda(\bm{s};t)$ vary continuously through
space.  The results from \citet{Lindgren+al:2011} show how to take
advantage of the connection between Gaussian Markov random fields of
graphs and stochastic partial differential equations in continuous
space. Some details of such models are given in Supplement~\ref{appx:spde} and 
the computational implications are discussed in Section~\ref{sec:SPDEcomputation}.

\subsection{Log-linear detection function models}
\label{sec:positivegx}

As noted in Section~\ref{sec:PPmodelsDist}, the hazard-rate model is
not a log-linear model, which means that estimating the parameters
does not directly fall under the latent Gaussian model framework of
the INLA estimation software~\citep{Rue+al:2009}. In contrast, the
half-normal detection model is $g_k(\mv{s})=\exp\left[-z_k(\mv{s})^2/(2\sigma^2_{g,k})\right]$, where $z_k(\bm{s})$ is the perpendicular distance from $\bm{s}$ to the
$k$th transect line segment, and $\sigma_{g,k}$ are scale parameters.
This can be written in log-linear form as
$\log\left[g_k(\mv{s})\right]=\beta_{g,k} z^*_k(\mv{s})$, where
$z^*_k(\mv{s})=-z_k(\mv{s})^2/2$, and $\beta_{g,k}=1/\sigma^2_{g,k}$.
To allow more flexibility within the log-linear framework we
introduce a semi-parametric piecewise quadratic model for the
logarithm of the detection function, based on a one-dimensional
version of the SPDE in the previous section.

For ease of presentation, first assume that the detection probability
is the same for all transects $k$, so that we can write
$G[z(\mv{s})]=-\log[g_k(\mv{s})]$. The prior distribution for $G(z)$
is then defined by the spline-like stochastic differential equation
\begin{align}
  \gamma\frac{\mmd^2 G(z)}{\mmd z^2} &= \mathcal{W}(t),\quad t\in[0,z_{\text{max}}]\subset\R, \label{eq:detsde}\\
  G(0)\;=\;0, &\;\;\;\;\;\;\;\;\;\;
  \left.\frac{\mmd G(z)}{\mmd z}\right|_{z=0}\;=\;0,\label{eq:bnd}
\end{align}
where $z_{\text{max}}$ is the maximal detection distance, $\gamma>0$
is a smoothness parameter, and $\mathcal{W}(t)$ is a white noise
process. The boundary constraints ensure that the detection
probability at distance $z=0$ is $1$, and that the probability is flat
near $z=0$.

\begin{figure}
  \subfigure[]{\includegraphics[width=0.3\linewidth]{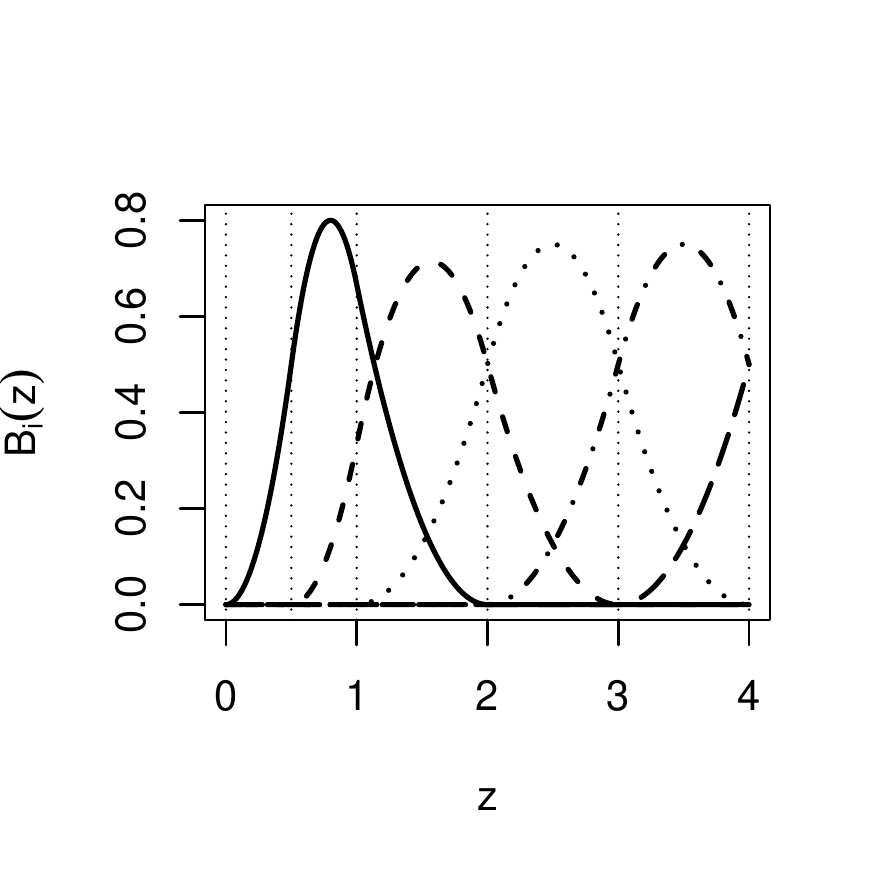}}
  \subfigure[]{\includegraphics[width=0.3\linewidth]{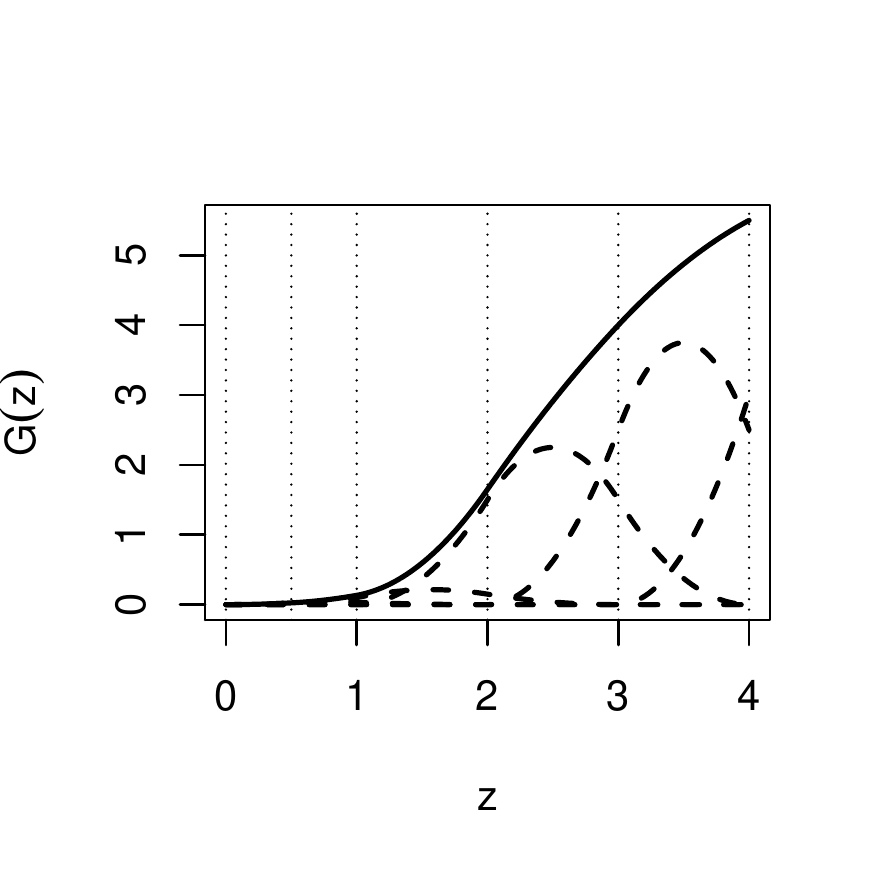}}
  \subfigure[]{\includegraphics[width=0.3\linewidth]{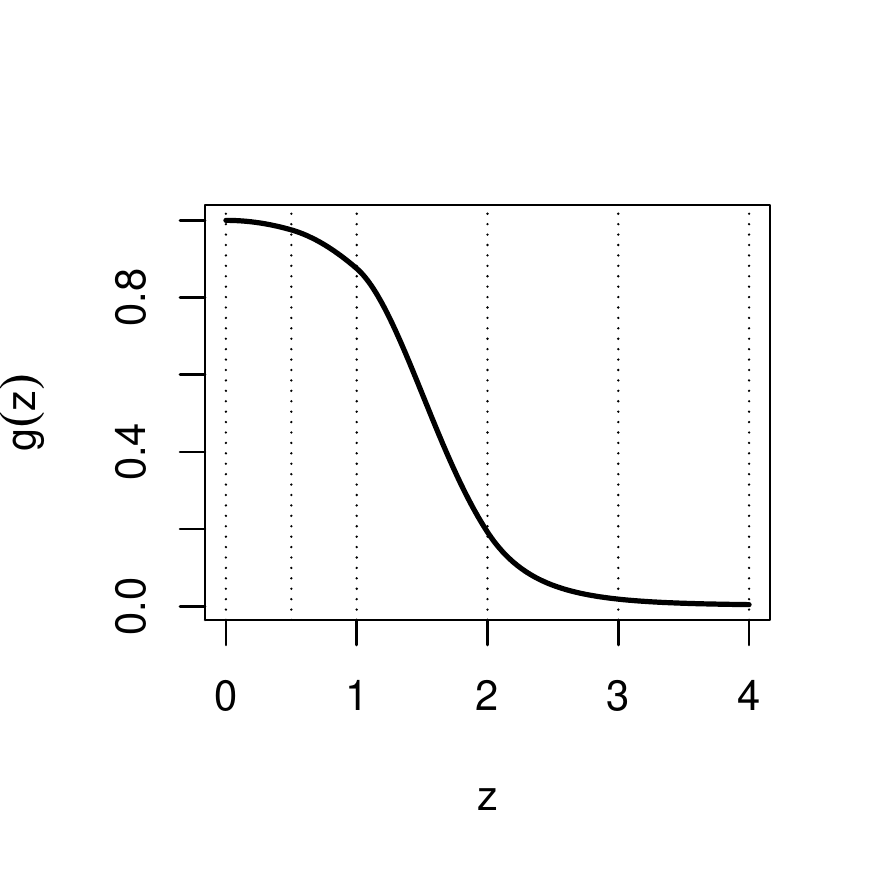}}
  \caption{Illustration of the semi-parametric detection function
    model: (a) B-spline basis functions $B_i(z)$, $i=1,\dots,5$, of
    order 2 that fulfill the boundary conditions~\eqref{eq:bnd}, (b) the weighted functions $\beta_iB_i(z)$
    (dashed), their sum $G(z)$ (solid), and (c) the detection
    probability $g(z)=\exp[-G(z)]$. The breakpoints are
    $(0,z_1,\dots,z_5)=(0,0.5,1,2,3,4)$ and the weights are
    $(\beta_1,\dots,\beta_5)=(0.05,0.3,3,5,6)$. Note that the
    detection function does not automatically have more irregular
    behavior where the breakpoints are close together, and that this
    is probabilistically regulated by the stochastic differential
    equation~\eqref{eq:detsde}.}
  \label{fig:detectionbasis}
\end{figure}

Let $(0,z_1,z_2,\dots,z_p)$ be breakpoints for piecewise quadratic
B-spline basis functions $B_i(z)$~\citep{Farin:2002}, such that
$z_p=z_{\text{max}}$, with the simplest choice of breakpoints being
$z_i=i z_{\text{max}} / p$, $i=0,1,\dots,p$.  The non-parametric model
for $G(z)$ can then be used to construct a finite dimensional model
\begin{align}
  G(z) &= \sum_{i=1}^p \beta_i B_i(z),
  \label{eq:semiparGz}
\end{align}
where the $p$ basis functions only include $B_i(\cdot)$ that fulfill
the boundary conditions \eqref{eq:bnd}, as shown in Fig~\ref{fig:detectionbasis}.
  The joint multivariate Gaussian prior distribution for
  $(\beta_1,\dots,\beta_p)\sim\pN\left(\mv{0},\mv{Q}_\beta^{-1}(\gamma)\right)$
  is constructed with the same finite element technique that will be
  used for the spatial SPDE in Section~\ref{sec:SPDEcomputation}.  For
  uniform breakpoint spacing, increasing $p$ will make this discrete
  model converge to the continuous domain model, but for finite $p$
  the model is effectively a piecewise quadratic semi-parametric model.

  Imposing a monotonicity constraint on $g(z)$ is possible by 
  replacing the basis functions for $G(z)$ with increasing basis
  functions, and mandating positivity of the $\beta_i$
  coefficients~\citep{Ramsay1988}.  However, because the latter is
  currently only implemented in INLA for independent $\beta_i$, this
  is restricted to small $p$, such as 2 or 3, since an independence
  prior would result in a non-smooth function for larger $p$.  An
  alternative that is feasible when sampling from the posterior
  distribution is to simply reject all non-increasing samples of
  $G(z)$.  If the data are informative, the smoothing from the prior
  can in practice be enough to yield monotonic estimates without
  including an explicit constraint.

  Relaxing the assumption that the detection function is the same for
  all transects is most easily done by adding further linear terms to
  $\log g_k(\mv{s})$ based on observed or constructed covariates that
  depend on $k$.

\section{Computational methods}
\label{sec:computation}
There are several practical considerations for evaluating the likelihoods and representing the random fields in such a way that large dense matrices can be avoided.  In Section~\ref{sec:SPDEcomputation} we give a brief overview of the essentials for translating stochastic PDEs into manageable Gaussian Markov random fields (GMRF), and Section~\ref{sec:numericalintegration} presents a numerical integration scheme for the point process likelihood.

\subsection{The stochastic partial differential equation approach}
\label{sec:SPDEcomputation}

The SPDE/GMRF approach works by replacing the continuous domain
stochastic PDE model with a finite dimensional Gaussian Markov random
field for basis function weights defined on a triangulation of the
domain of interest, such that the sparse precision matrix leads to a
good approximation of the continuous space SPDE solutions.  Given a
triangulation mesh (see the right panel of Fig~\ref{fig:effortsight} for the triangulation
used for the ETP survey area), \citet{Lindgren+al:2011} define a
finite element representation \citep{Brenner+Scott:2007} of $\xi$ from
\eqref{eq:SPDE},
\begin{align}
\label{eq:fem}
\xi(\bm{s}) = \sum_{j=1}^m w_j \phi_j(\bm{s}),
\end{align}
where
$w_1, \dots, w_m$ are stochastic weights, and $\phi_j$, $j=1, \dots, m$, are deterministic piecewise linear basis functions defined for each node on the mesh: $\phi_j$ equals $1$ at mesh node $j$ and $0$ in all the other mesh nodes. The weight vector $\bm{w}\equiv(w_1, \dots, w_m)^\top$ is a GMRF with its Markovian properties defined by the mesh structure. It follows that $\bm{w}$ determines the stochastic properties of (\ref{eq:fem}) and $\bm{w}$ is chosen in a way that the distribution of (\ref{eq:fem}) approximates the distribution of the solution to the SPDE (\ref{eq:SPDE}). As shown by \citet{Lindgren+al:2011}, for the SPDE in \eqref{eq:SPDE},
the resulting weight distribution is
$\bm{w}\sim\pN(\bm{0},\bm{Q}(\tau,\kappa)^{-1})$, where the sparse
precision matrix $\bm{Q}(\tau,\kappa)$ is a polynomial in the
parameters $\tau$ and $\kappa$, and is obtained through finite element
calculations.

The practical implication of this construction is that instead of
directly using the covariances from \eqref{eq:materncov}, which
results in dense covariance matrices and high computational cost,
$\mathcal{O}(m^3)$, the SPDE/GMRF approach links the continuous and
discrete domains in such a way that the computational cost is reduced
to $\mathcal{O}(m^{1.5})$.  The computational advantages of GMRFs
\citep{Rue+Held:2005} is strengthened by using INLA for Bayesian
inference \citep{Rue+al:2009}.  For the case of fully observed
log-Gaussian Cox point processes, the in-depth analysis by
\citet{Simpson+al:2015} of the combined approximation errors induced
by the basis function expansion in combination with the likelihood
approximation error shows that the resulting approximate posterior
distribution is close to the true posterior distribution.  Since the
integration scheme in the following section is constructed in the same
way, we do not include a detailed approximation error analysis here,
but note that the SPDE/GMRF approximation is likely to be the largest
source of approximation error. Point patterns are relatively
uninformative about the latent intensity, which has the practical
effect that the realizations of the fields in the posterior
distribution are typically smoother than in directly observed process
problems.  Hence, the approximation error is very small as long as the
triangle mesh edges are short compared with the spatial scales of the
covariates and of the point pattern intensity variability.

\subsection{Numerical point process likelihood evaluation}
\label{sec:numericalintegration}

Combining the general distance sampling point pattern likelihood
\eqref{eq:likelihood} with the log-linear model structure for
$\Lambda(\bm{s};t)$ from \eqref{eq:logLambda} results in a
log-likelihood for the observed point pattern,
\begin{align}
\log \pi(\bm{Y}\mid\lambda,g)=& \sum_{i=1}^{N_{\bm{Y}}} \left\{\bm{x}(\bm{s}_i, t_i) ^\top \bm{\beta} +\xi(\bm{s}_i,t_i)  -\log[g_{k(t_i)}(\bm{s}_i)]\right\} \non\\
&- \sum_{k=1}^K\int_{\CC_k}\exp\left\{\bm{x}(\bm{u},t_k)^\top \bm{\beta} +\xi(\bm{u},t_k)  -\log[g_{k}(\bm{u})]\right\} \md\bm{u}
+ \sum_{k=1}^K |\CC_k|
,
\label{eq:filterPPP}
\end{align}
where the first term evaluates the log-intensity at the observed
locations, and the second term integrates the intensity over the
sampled transect segments. The log-likelihood (\ref{eq:filterPPP}) is
in general analytically intractable as it requires integrals of the
exponential of a random field. Therefore, we use numerical integration
to evaluate (\ref{eq:filterPPP}), and the remainder of this section
describes an integration scheme to approximate the integrals
efficiently.  As noted in Section~\ref{sec:model}, we assume that the
covariates $x(\bm{s},t)$ are expressed using the same piecewise linear
basis functions as $\xi$.  For cases where a covariate has a much
finer resolution than the one needed for $\xi$, the efficient
integration scheme developed here is not appropriate, and further
research is needed to develop an integration method that can deal with
that without incurring a high computational cost.

For distance sampling surveys, transect areas describe subsets of the earth's surface. The most natural representation of transect areas would therefore be subsets $\CC_k\subseteq\mathbb{S}^2$ of the sphere, leading to surface integration in the Poisson process likelihood. However, the small
scale at which earthbound observers are capable of probing their
environment lends itself to easily justifiable simplifications of
the numerical integration. Apart from environmental conditions such as the
weather, the curvature of the earth puts an upper bound on the
distance at which an observer with a given elevation can actually
detect an animal. We therefore approximate the surface integrals over $\CC_k\subset\mathbb{S}^2$ by integrals over $\wt{\CC}_k\subset\R^2$,
\begin{align*}
I_{\CC_k} &= \int_{\CC_k}\li(\bm{u};t_k) g_k(\bm{u}) \md\CC_k(\bm{u})\;=\; \int_{\wt{\CC}_k}\li(\bm{u}_k(l,z);t_k) g_k(\bm{u}_k(l,z))  \left\| \frac{\partial \bm{u}_k}{\partial l} \times \frac{\partial \bm{u}_k}{\partial z} \right\| \md l \md z \\
&\approx \int_{\wt{\CC}_k} \li(\bm{u}_k(l, z);t_k) g_k(\bm{u}_k(l,z)) \md l \md z,
\end{align*}
where we use a transect-specific parameterization $\bm{u}_k$ at
coordinate $l$ along and distance $z$ to the transect line,
respectively. If $R$ is the radius of the earth, then the Jacobian is $\cos(z/R)$, which gives an approximation error of a factor less than $5\cdot 10^{-6}$ even in the extreme and unrealistic case of an observer at $31$ metres above a calm sea looking at the horizon $20$ kilometres away.

\begin{figure}
\subfigure[]{\includegraphics[width=0.3\linewidth]{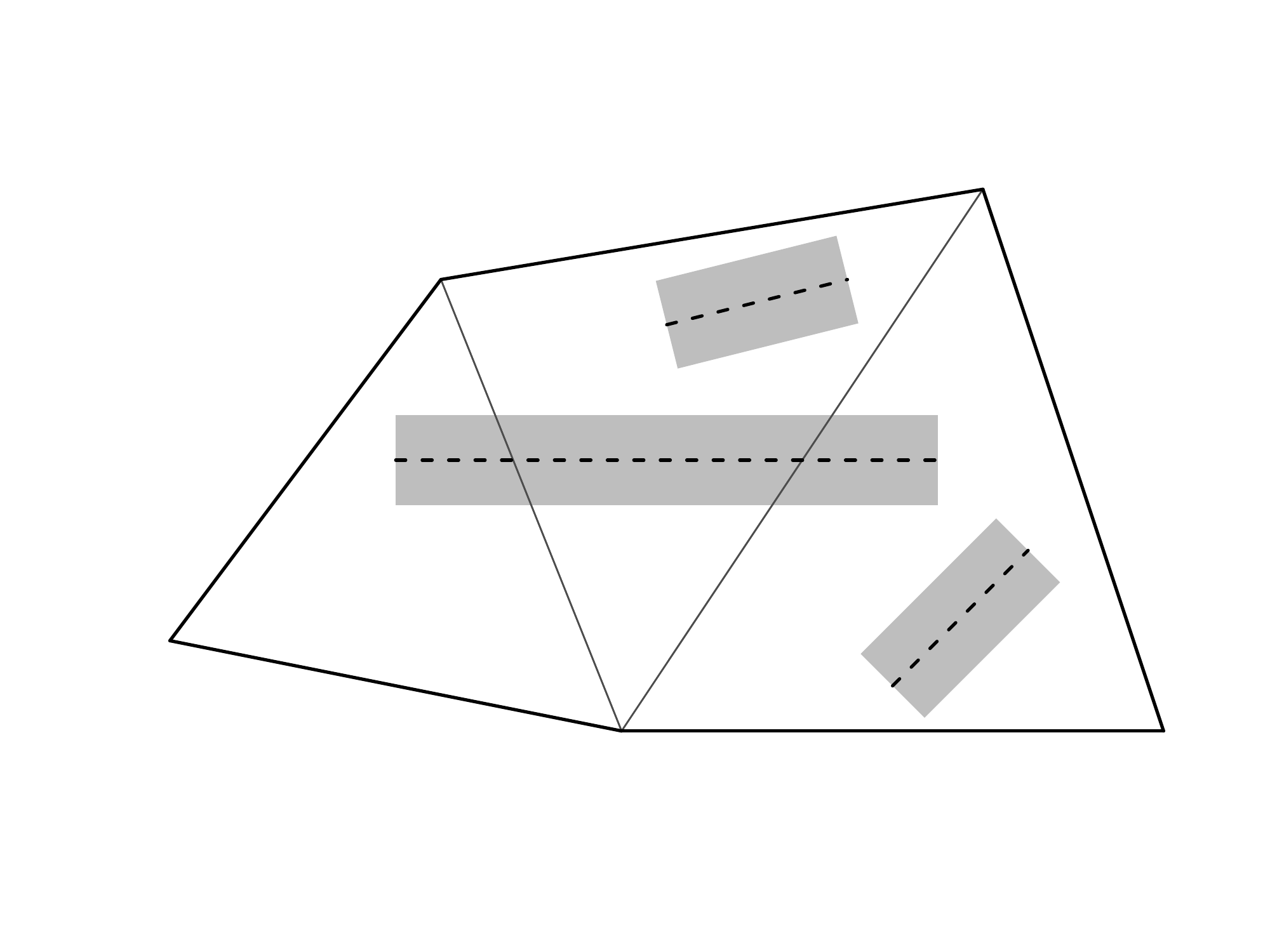}}
\subfigure[]{\includegraphics[width=0.3\linewidth]{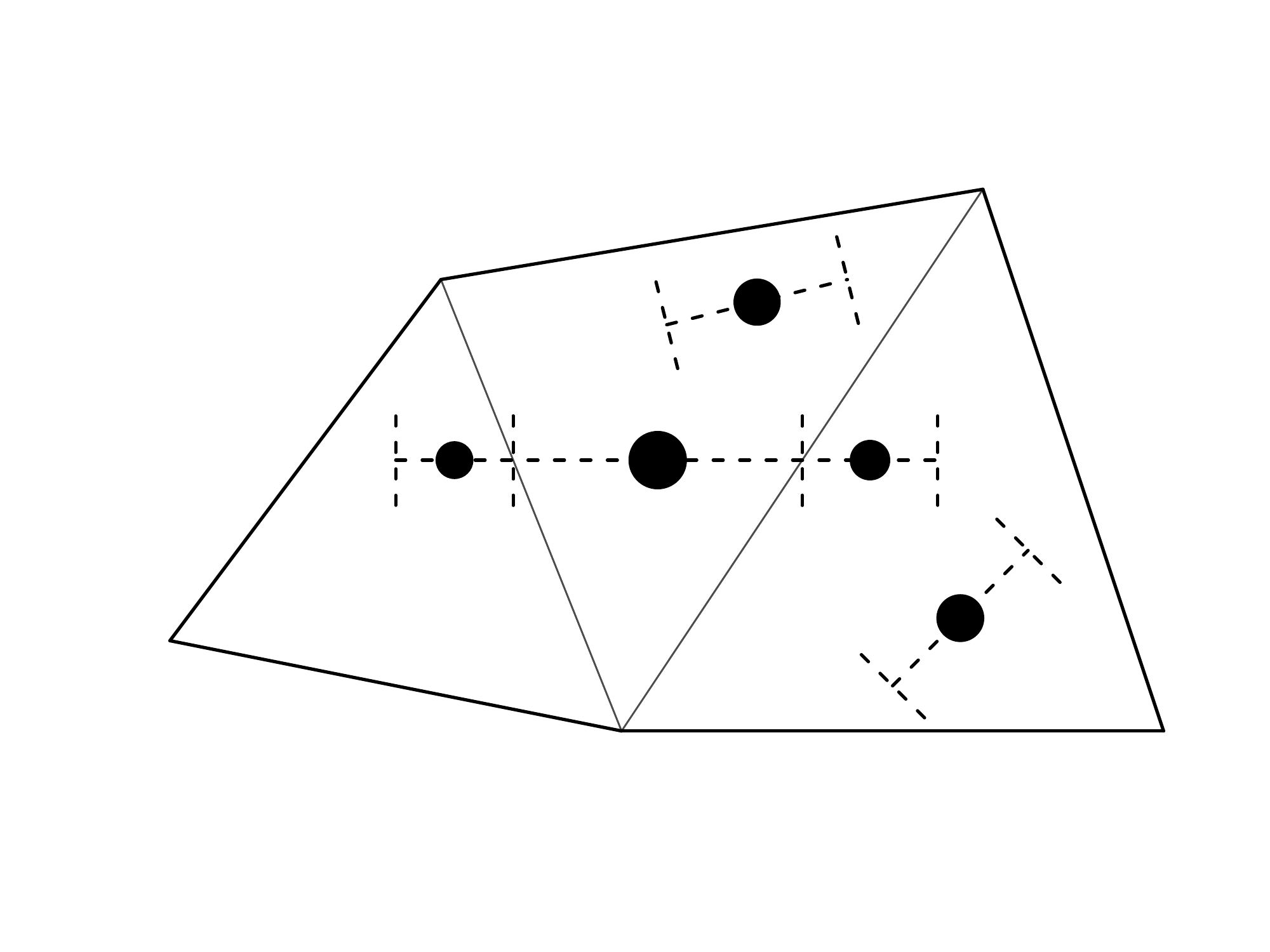}}
\subfigure[]{\includegraphics[width=0.3\linewidth]{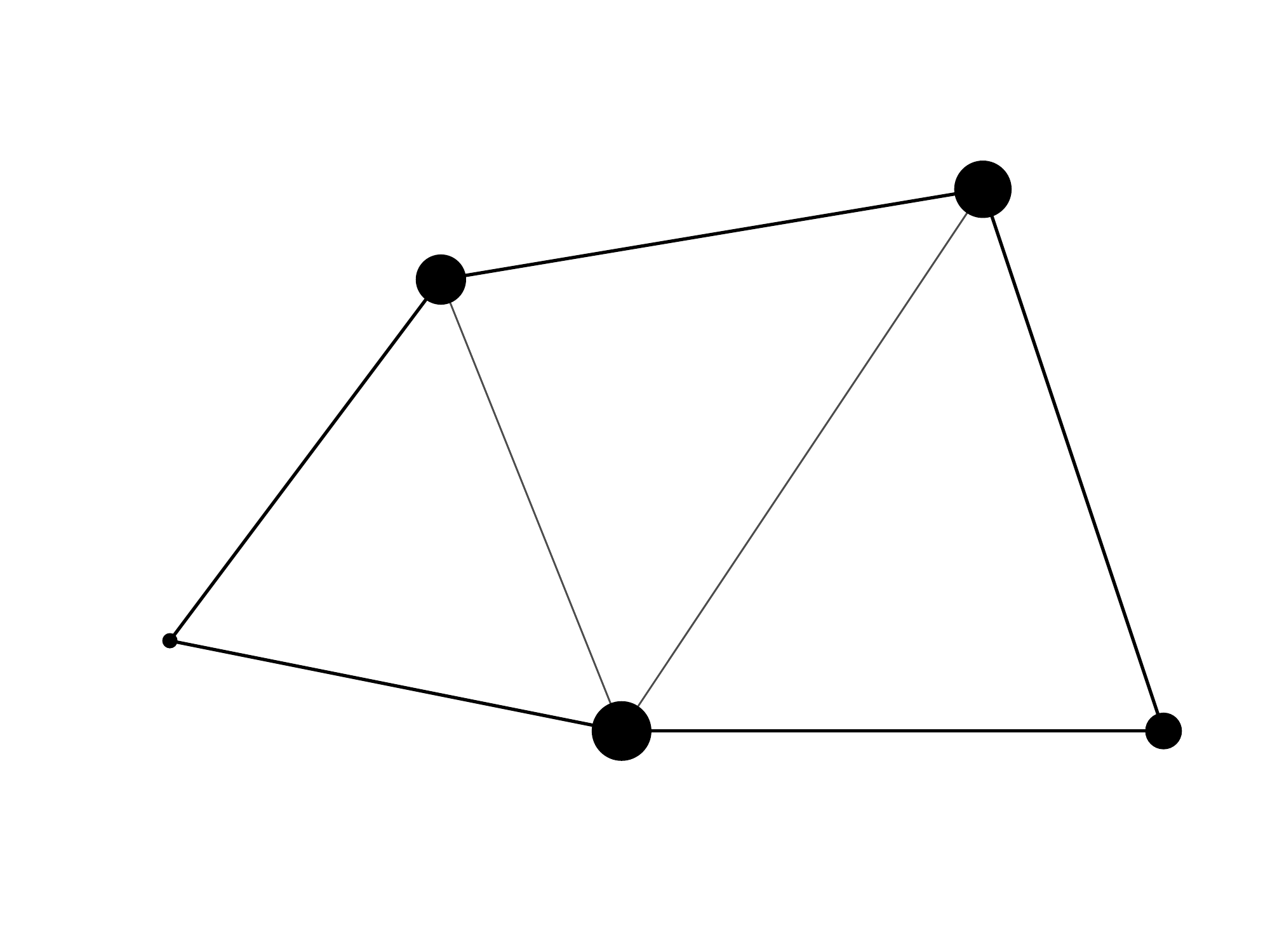}}
\caption{Integration scheme for transect lines. Panel (a) shows three triangles of a mesh, and three transects. When the width of the transects is small compared to the size of the triangles, and thus the slope of the intensity has small variability perpendicular to each segment, the areal integration can be reduced to an integration along the observer's trajectory weighted by the width of the transect. Trajectories that reside in multiple triangles can be split at the triangle edges such that their midpoints serve as integration points, assuming linear intensity within the triangles (b). Under the same assumption, integration points from each triangle are accumulated into re-weighted points at the mesh vertices (c). The areas of the filled circles are proportional to the integration weights.}
\label{fig:integration_scheme}
\end{figure}

Another fact that we can utilize is that the detection function $g$ does not depend on
the position of the observer along the line but only on the distance of an
observation from the line. Similarly, if the transect line is narrow compared to the
spatial rate of change in the intensity function, we can substitute
the evaluation of $\lambda$ by an evaluation at the center of the transect line, $\wh{z}=0$ (see Fig~\ref{fig:integration_scheme}). That is,
\begin{align*}
I_{\CC_k} &\approx \int_{\wt{\CC}_k} \li(\bm{u}_k(l,\wh{z});t_k) g_k(\bm{u}_k(\wh{l},z)) \md l \md z,
\end{align*}
together with an arbitrary coordinate $\wh{l}$ along the transect
line.
\begin{remark}
  In a standalone implementation, the integral could be written as a
  product of two one-dimensional integrals, and even evaluated exactly
  due to the log-linearity of the model.  Unfortunately, the resulting
  structure cannot be expressed using only evaluations of products of
  $\lambda$ and $g$, which is a requirement imposed by the internal
  structure of the INLA implementation, so we do not use that approach
  here.
\end{remark}

We can also make use of the fact that $\li$ lives on a mesh. If
the mesh triangles are small enough, the log-linear function
$\li(\cdot)$ is approximately linear within each triangle. By
splitting a transect line $\CC_k$ into segments $\CC_{k,j},\,j \in 1,\dots,J$, each of
which resides in a single triangle (see Fig~\ref{fig:integration_scheme}), we obtain a Gaussian quadrature method
of order one,
\begin{align*}
I_{\CC_k} &\approx \sum_{j=1}^J\int_{\wt C_{k,j}} \li(\bm{u}_{kj}(l,\wh{z});t_k) g_k(\bm{u}_{kj}(\wh{l},z)) \md l \md z 
\;\approx\; \sum_{j=1}^J w_{k,j}\int_{z} \li(\bm{u}_{kj}(l_{k,j},\wh{z});t_k) g_k(\bm{u}_{kj}(\wh{l},z)) \md z.
\end{align*}
Here, $l_{k,j}$ is half of segment $j$'s length $w_{k,j}$. The integration over the distance parameter can now be approximated by a quadrature rule with an equidistant scheme, so that
\begin{align*}
I_{\CC_k} &\approx \sum_{j=1}^J \sum_{r=1}^R \wt{w}_{k,j} \li(\bm{u}_{kj}(l_{k,j},\wh{z});t_k) g_k(\bm{u}_{kj}(\wh{l},z_r)),
\end{align*}
where $\wt{w}_{k,j} = \frac{2z_{\textrm{max}}}{R}w_{k,j}$ with maximal detection distance  $z_{\textrm{max}}$, and we substitute $\wh{l} = l_{k,j}$.
We can then write
\begin{align*}
I_{\CC_k} &\approx \sum_{j=1}^J \sum_{r=1}^R \wt{w}_{k,j} \li(\wt{\bm{u}}_{k,j,r};t_k) g_k(\bm{u}_{k,j,r}),
\end{align*}
where $\bm{u}_{k,j,r}$ are points on the perpendicular line
through the midpoint of transect k's segment $j$, and
$\wt{\bm{u}}_{k,j,r}$ is the midpoint of each subsegment line.

As a last step we can again make use of the assumption that the
function we are integrating over is approximately linear within a
given triangle. It is straightforward to show (see
Supplement~\ref{appx:integration_projection}) that this means that each
integration point can be expressed by an evaluation of the function at
the triangle vertices weighted by the within-triangle Barycentric
coordinates of the original point~\citep{Farin:2002}. We can therefore
summarize integration points that reside in the same triangle and
share a common time coordinate $t_k$ to such evaluations at the mesh
vertices, illustrated in Fig~\ref{fig:integration_scheme}.  This can,
depending on the problem structure, lead to a significant reduction in
the respective computational workload.

The approximation error from treating the log-linear function within
each triangle as linear can be reduced by subdividing each triangle
into four.  However, evaluating the function at the mid-points of the
original triangle edges as well as the original vertices leads to an
increase in the computational cost of at least a factor of four, since
the number of edges is approximately three times the number of
vertices in the mesh.

\section{Estimating blue whale density from the Eastern Tropical Pacific surveys}
\label{sec:casestudy}

\subsection{The ETP surveys}

We use the above methods to predict the blue whale group density over the ETP survey area for the each of the survey years, and to study the effect of sea surface temperature (SST) on the blue whale group density. The function of interest for density estimation is the intensity of the point process before thinning, denoted $\lambda(\bm{s}, t)$ in \eqref{eq:logLambda}, which we refer to as the group density.

These data have been analyzed before: \citet{Forney+al:2012} used GAMs to estimate encounter rate, with a two-stage estimation approach and gridded data (counts of detections within small segments of transects). \citet{Pardo+al:2015} also used gridded data, modelling log density in each grid cell as a polynomial function of the absolute dynamic topography, a spatially referenced variable that indicates vertical transport of nutrients and thus productivity. While they included a random component in their density model, it had no spatial structure, assuming independent residuals among grid cells.  They estimated all model parameters simultaneously in a hierarchical Bayesian framework. Two key differences between their and our model structures are that we use the ungridded data (i.e. the point locations of each detection rather than counts in user-defined grid cells) in our analysis, and we use a \textit{spatially structered} Gaussian random field to capture spatial variation in density that is not explained by the observed explanatory variable.

Because the blue whale group size is small (mean of 1.8 and standard deviation 2.1) and the size is easily established, it is realistic to treat the group size of blue whales as known without error \citep{Gerodette+al:2002}. We assume that the group size does not affect the detection probability. The detection of cetaceans on ship surveys also depends on wind conditions, but this is less important for blue whales because of their large body size and conspicuous blows. Therefore, we assume the detection probability depends only on the perpendicular distance for the blue whales in the ETP survey. In our analysis, we truncate the data at perpendicular distance $w=6$~km. We also assume that distances were observed without error for each detected animal group and we fit a semi-parametric model given by \eqref{eq:semiparGz} to estimate the detection function.

To build a spatio-temporal model using the SPDE approach described in Section~\ref{sec:SPDEcomputation}, we start by constructing a mesh for the ETP survey as shown in Fig~\ref{fig:effortsight}. The ETP survey is bounded partially by the coastline and partially by the red line of Fig~\ref{fig:effortsight}. We use a simplified representation of the actual coastline as the mesh boundary to incorporate the boundary effect because a physical boundary such as the coastline has a strong effect -- in this case, there will be no blue whales on land. We use a simplified representation of the coastline, because the actual coastline is too `angular' and hence problematic for the SPDE approach \citep{Lindgren+Rue:2015}. Meanwhile, we extend the boundary further in the northwest and south directions in the Eastern Tropical Pacific, to exclude the boundary effect for the part of the survey area that is bounded by the red line in Fig~\ref{fig:effortsight}. Given the low sighting rate of the blue whales, there is little information contained in the data to fit a complicated spatio-temporal stochastic process for the random field, such as the AR(1) temporal process used by \citet{Cameletti+al:2011}; even the simpler version of the spatio-temporal process with replicated spatial field over time is not feasible. 

We consider three models, all with Gaussian random fields in space alone. Model 0 has latitude and longitude fixed effects but ignores SST. Model 1 has a temporal SST fixed effect together with spatial residual SST fixed effects for each year. Model 2 has a temporal SST fixed effect and a spatial SST fixed effect together with spatio-temporal residual SST fixed effects.

\subsection{Incorporating a spatio-temporal environmental covariate: sea surface temperature (SST)}\label{sec:centeringSST}

Based on the Simple Ocean Data Assimilation (SODA) model ({\small\url{http://apdrc.soest.hawaii.edu/datadoc/soda_2.2.4.php}}), the SST data are available on a fine grid over the ETP survey area on a monthly scale between 1986 and 2007. First,  within each year, SST is averaged over the months July to December, during which the survey was conducted. Second, these temporally averaged SST values are spatially smooth, and can be projected onto the mesh of the survey area with only minor loss of fine-scale information. Piecewise linear interpolation is used to calculate the SST for any given location and year, denoted by $\sst(\bm{s}, t)$.

\begin{figure}[t]
	\centering
	\includegraphics[width=0.45\textwidth]{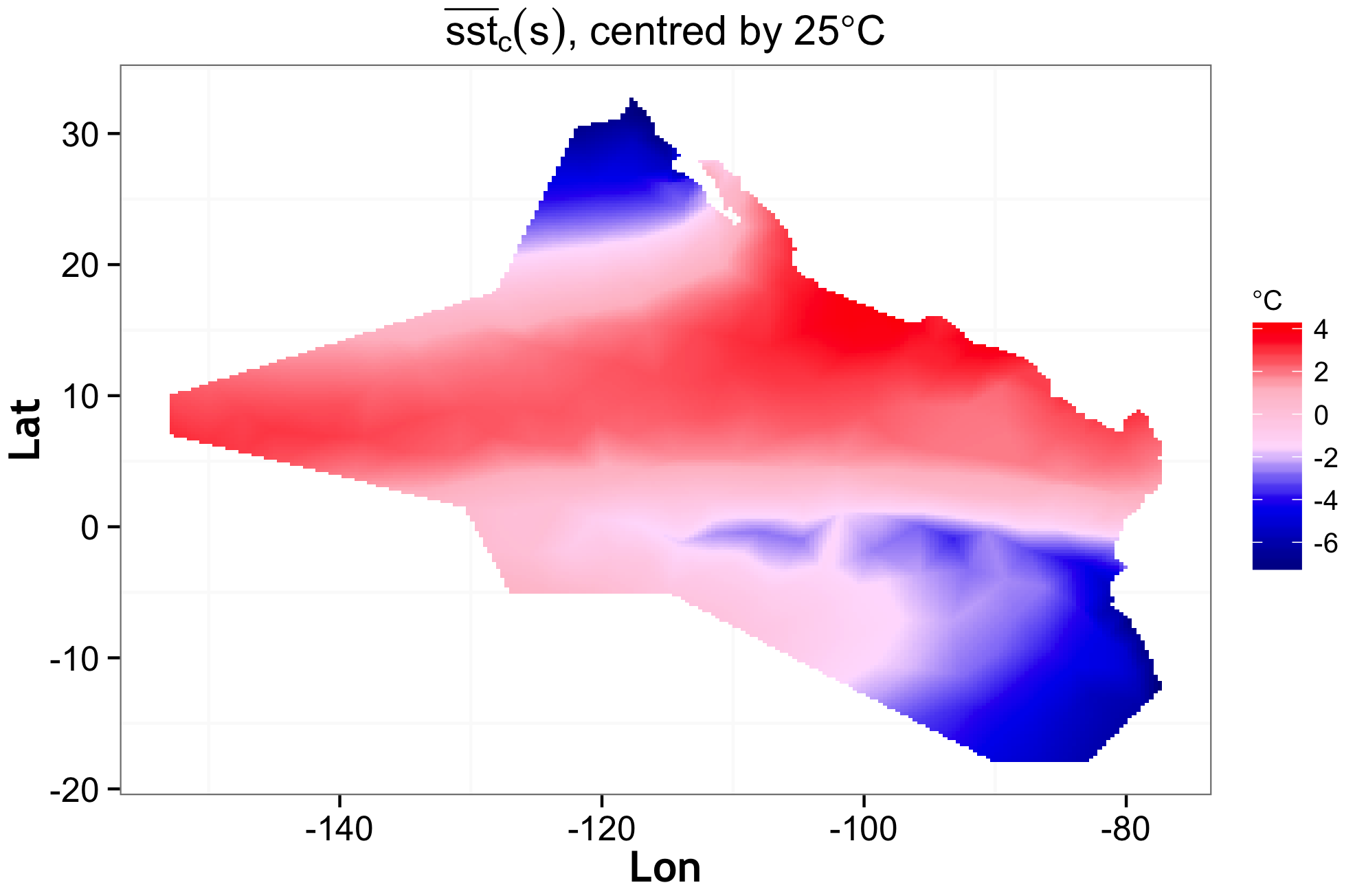}\hfil
	\includegraphics[width=0.35\textwidth]{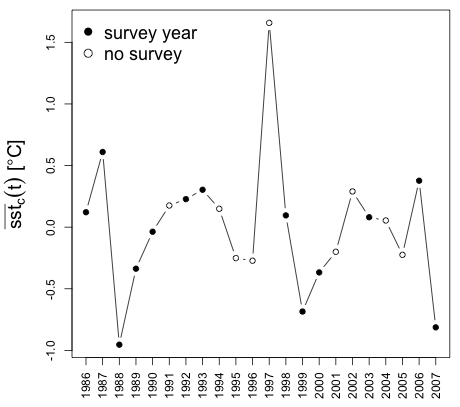}
	\caption{Sea surface temperature. The centered time-average temprature $\overline{\sst}_c(\bm{s})$ is shown in the left panel, while the  centered space-averaged temperature $\overline{\sst}_c(t)$ is shown in the right panel. Solid circles represent survey years and the empty circles non-survey years. The SST in 1997 is extreme relative to all survey years.  }
	\label{fig:barSSTcYear}
\end{figure}

Fig~\ref{fig:barSSTcYear} shows the centered SST averaged over time and the centered SST averaged over space. There is both spatial and inter-year variation in SST, and we use hierarchical centering to separate the annual and spatial effects of SST. 

Hierarchical centering is a commonly used technique in multilevel modeling \citep{Kreft+al:1995}, and we consider two different centering schemes for SST here. Model 0 does not incorporate SST, Model 1 incorporates SST using within-year centering, and Model 2 uses space-time centering. They all have the same SPDE specification of the latent spatial Gaussian random field.

\begin{enumerate}
\item Model 1: within-year centering. This model has two SST components, the spatially averaged SST for each year, and the spatial SST patterns centered within each year. Let $\Omega$ denote the bounded ETP survey area in Fig~\ref{fig:effortsight}. We use $\overline{\sst}_c(t)$ to denote the SST averaged over the ETP survey area for year $t$ after centering,

\begin{align}
\overline{\sst}_{c}(t)&=\frac{1}{|\Omega|}\int_{\Omega} \sst(\bm{s}, t) \md\bm{s} -\overline{\sst},
\label{eq:SSTcYr}
\end{align}
where $\overline{\sst}$ denotes the overall average of SST, $\overline{\sst} =\int_{\Omega\times\mathbb{T}} \sst(\bm{s}, t) \md\bm{s} \md t\big/(|\Omega|\times |\mathbb{T}|)$, with $\mathbb{T}$ denoting the set of survey years. Then the SST centered within year $t$ for location $\bm{s}$, $\sst_{cwy}(\bm{s}, t)$, is defined as
\begin{align}
\sst_{cwy}(\bm{s}, t) &= \sst(\bm{s}, t) -  \overline{\sst}_{c}(t).\label{eq:SSTcwy}
\end{align}

\item Model 2: space-time centering. This model separates the spatial and temporal patters from a spatio-temporal interaction and has three SST components. These are the $\overline{\sst}_c(t)$ given by \eqref{eq:SSTcYr}, the SST averaged over years for each location, $\overline{\sst}_c(\bm{s})$ given by \eqref{eq:SSTcLoc}, and the SST residuals, $\sst_{res}(\bm{s}, t) $ given by \eqref{eq:SSTres},

\begin{align}
\overline{\sst}_c(\bm{s}) &= \frac{1}{ |\mathbb{T}|}\int_{\mathbb{T}} \sst(\bm{s}, t) \md t  - \overline{\sst}, \label{eq:SSTcLoc}\\
\sst_{res}(\bm{s}, t) &= \sst(\bm{s}, t) - \overline{\sst}_c(t) -\overline{\sst}_c(\bm{s}) - \overline{\sst}.\label{eq:SSTres}
\end{align}
$\overline{\sst}_c(t)$ indicates whether a year is relatively warm or cold after averaging over the survey area, and similarly, $\overline{\sst}_c(\bm{s})$ indicates whether a location is relatively warm or cold after averaging over the survey years. The SST residual, $\sst_{res}(\bm{s}, t)$, contains information about the interaction between temporal pattern and the spatial average of SST. 

\end{enumerate}

Plots of raw and centered SST used in the analysis are given in Supplement~\ref{appx:plotSST}.

La Ni\~{n}a conditions are characterized by a band of cooler waters in 1988, 1999 and 2007, and El Ni\~{n}o conditions by a much wider band of warm ocean water in 1997 \citep[1987 is a moderately strong El Ni\~{n}o year according to the scale by][]{Wolter+Timlin:2011}. Centering SST strongly captures the El Ni\~{n}o/La Ni\~{n}a oscillations that occur at irregular intervals in the ETP survey area (see Supplement~\ref{appx:plotSST} for more detail). The temporal effect of SST after centering using (\ref{eq:SSTcYr}) (see Fig~\ref{fig:barSSTcYear}) correctly reflects the La Ni\~{n}a conditions in 1988 and 2007, and strongly highlights the El Ni\~{n}o year 1997 as an outlier. Unfortunately, no survey was conducted in 1997. Given the time series of $\overline{\sst}_{c}(t)$ for the survey years and non-survey years in the right panel of  Fig~\ref{fig:barSSTcYear},  it is obviously problematic to predict for 1997 using a model fitted on the data from the survey years, which are represented by the filled circles in Fig~\ref{fig:barSSTcYear}. Therefore, we make predictions for all years except 1997.

\subsection{Results}\label{sec:ETPresults}
Table~\ref{tab:betasAllmodel} summarizes the posterior density of the regression coefficients for each model. 

\subsubsection{The effects of longitude and latitude}
Model 0 contains only longitude and latitude as covariates. The 95\% posterior credible intervals for the regression coefficients from this model both include zero with medians very close to zero. This suggests that there is no large-scale log-linear spatial effect that can be explained by longitude and latitude. This interpretation is supported by the results from models that include SST. Specifically, when we add longitude and latitude to Models 1 and 2, the 95\% posterior credible intervals of the longitude and latitude regression parameters still include zero. We therefore exclude longitude and latitude, and henceforth consider only Models 1 and 2.


\begin{table}[t]
\centering
\caption{The posterior estimates for the fixed-effects coefficients for each model.}
\label{tab:betasAllmodel}
\begin{tabular}{lcccccc}
&& & &\multicolumn{3}{ c }{Quantile} \\
  \hline
Model &Parameter & Mean & Std.dev. & 2.5\%  & 50\%& 97.5\%  \\
  \hline
   \multirow{3}{*}{Model 0}&
$\displaystyle{\beta_0}$ &-12.29& 2.29 & -18.04&-11.99& -8.56\\ \addlinespace[0.56ex]
  & $\beta_{lon}$ &  0.10  & 0.07 & -0.05 &0.10 &0.26  \\  \addlinespace[0.56ex]
  &$\beta_{lat}$ & 0.01& 0.09 & -0.22 & 0.02& 0.16  \\
 \addlinespace[0.56ex]  \hdashline  \addlinespace[0.56ex]
  \multirow{3}{*}{Model 1}
  &$\beta_0$                   &-4.58& 3.04 & -11.00& -4.44& 1.06  \\ \addlinespace[0.56ex]
  &  $\beta_{\overline{\sst}_c(t)}$ & 0.79  & 0.21 &0.38 &0.78 & 1.20  \\
  \addlinespace[0.56ex]
  &  $\beta_{\overline{\sst}_{cwy}(\bm{s}, t)}$ & -0.28 & 0.10 & -0.48&-0.28  & -0.07  \\ \addlinespace[0.56ex]
\addlinespace[0.56ex] \hdashline \addlinespace[0.56ex]
    \multirow{3}{*}{Model 2}
     &$\beta_0$                       &-11.85  & 2.24 & -17.34&-11.60& -7.85  \\
     \addlinespace[0.56ex]
     &$\beta_{\overline{\sst}_c(t)}$  & 0.73  & 0.21 & 0.32& 0.73 & 1.16  \\ \addlinespace[0.56ex]
    &$\beta_{\overline{\sst}_c(\bm{s})}$ & -0.60& 0.14 & -0.88 &-0.60 & -0.34  \\ \addlinespace[0.56ex]
    &$\beta_{\overline{\sst}_{res}}$ & 0.22& 0.17 & -0.10 &0.22  & 0.55  \\ \addlinespace[0.56ex]
 \hline
\end{tabular}
\end{table}


\subsubsection{The effects of SST}
From \eqref{eq:SSTcwy}, \eqref{eq:SSTcLoc} and \eqref{eq:SSTres}, we have $\overline{\sst}_{cwy}(\bm{s}, t) =\overline{\sst}_{c}(\bm{s}) + \overline{\sst}_{res}(\bm{s}, t)$, so that $\beta_{\overline{\sst}_{cwy}(\bm{s}, t)}$ in Model 1 amounts to combining $\beta_{\overline{\sst}_c(\bm{s})}$ and $\beta_{\overline{\sst}_{res}}$ in a single parameter. The negative posterior median and 95\% credible interval (2.5\% to 97.5\% quantiles) of $\beta_{\overline{\sst}_c(\bm{s})}$ and of $\beta_{\overline{\sst}_{cwy}(\bm{s}, t)}$ indicate that locations that are colder on average over the years are expected to have more blue whale groups than locations that are warmer on average, while the opposite sign of the spatio-temporal interaction $\beta_{\overline{\sst}_{res}}$ indicates that this effect is weaker at locations with higher temperature in a given year than the across-year average temperature at the location.

\begin{figure}[h!]
\centering
\includegraphics[width=0.75\textwidth]{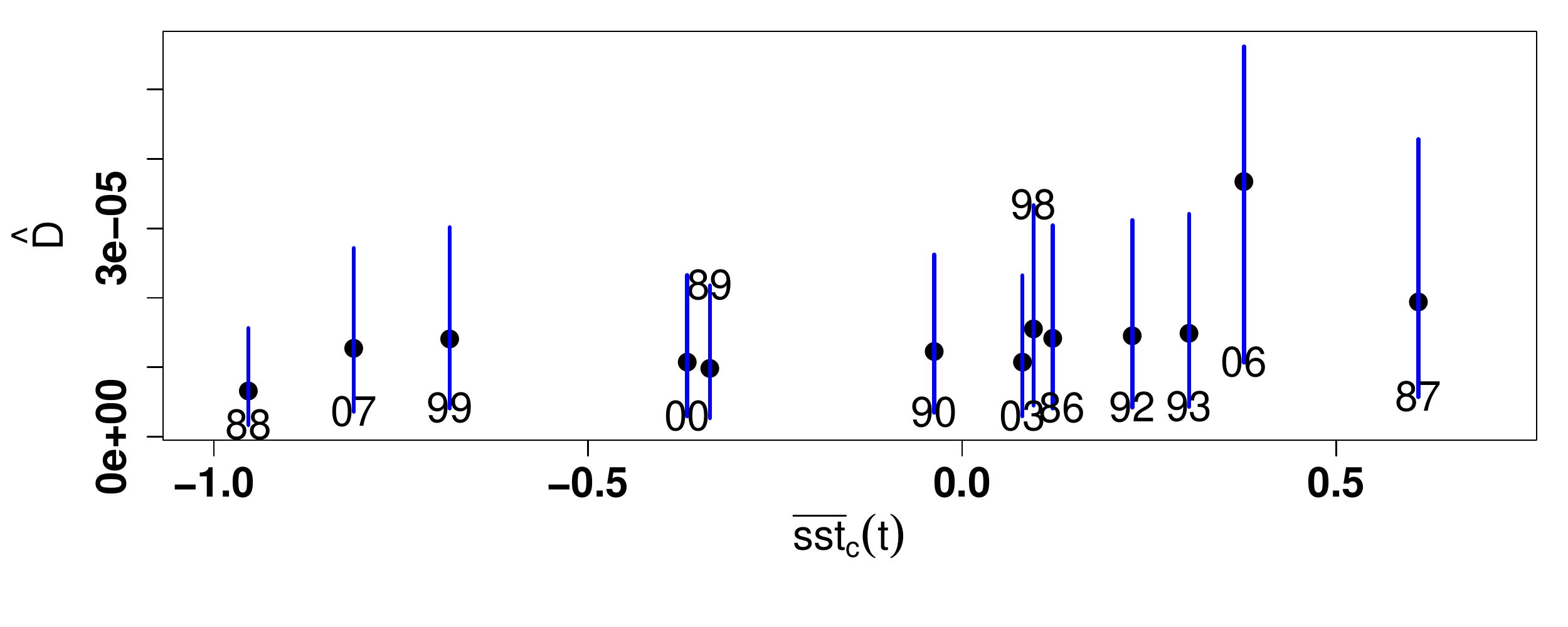}
\caption{The predicted number of blue whale groups per unit area ($\hat{D}$) from Model 2, together with 95\% credible intervals against centered mean annual temperature $\overline{sst}_c(t)$. Numbers indicate the year in question.}
\label{fig:DvsT}
\end{figure}

The posterior median estimates of $\beta_{\overline{\sst}_c(t)}$ are similar for Models 1 and 2, indicating that the effect of warmer average temperature in a year, conditional on the spatial effect and random field, is to increase density. The ETP survey design is not balanced in that it does not have survey effort in every year along each transect that was surveyed in any year and as a result we need to be a bit cautious about interpretating parameters. To investigate the effect of annual mean temperature, we therefore also considered the posterior distribution of the predicted number of blue whale groups per unit area. This is shown in Figure~\ref{fig:DvsT}. While this plot is consistent with the estimates of $\beta_{\overline{\sst}_c(t)}$ from Models 1 and 2, it is also consistent with an hypothesis of no change in average density across the years, as a horizontal line falls well within the 95\% credible intervals of all estimates. The other notable feature of the plot is the unusually high estimated density for the second-warmest year, 2006. The reasons for this are unclear.

\subsubsection{Posterior median density and its relative uncertainty}

\begin{figure}
	\centering
	\includegraphics[width=.8\linewidth]{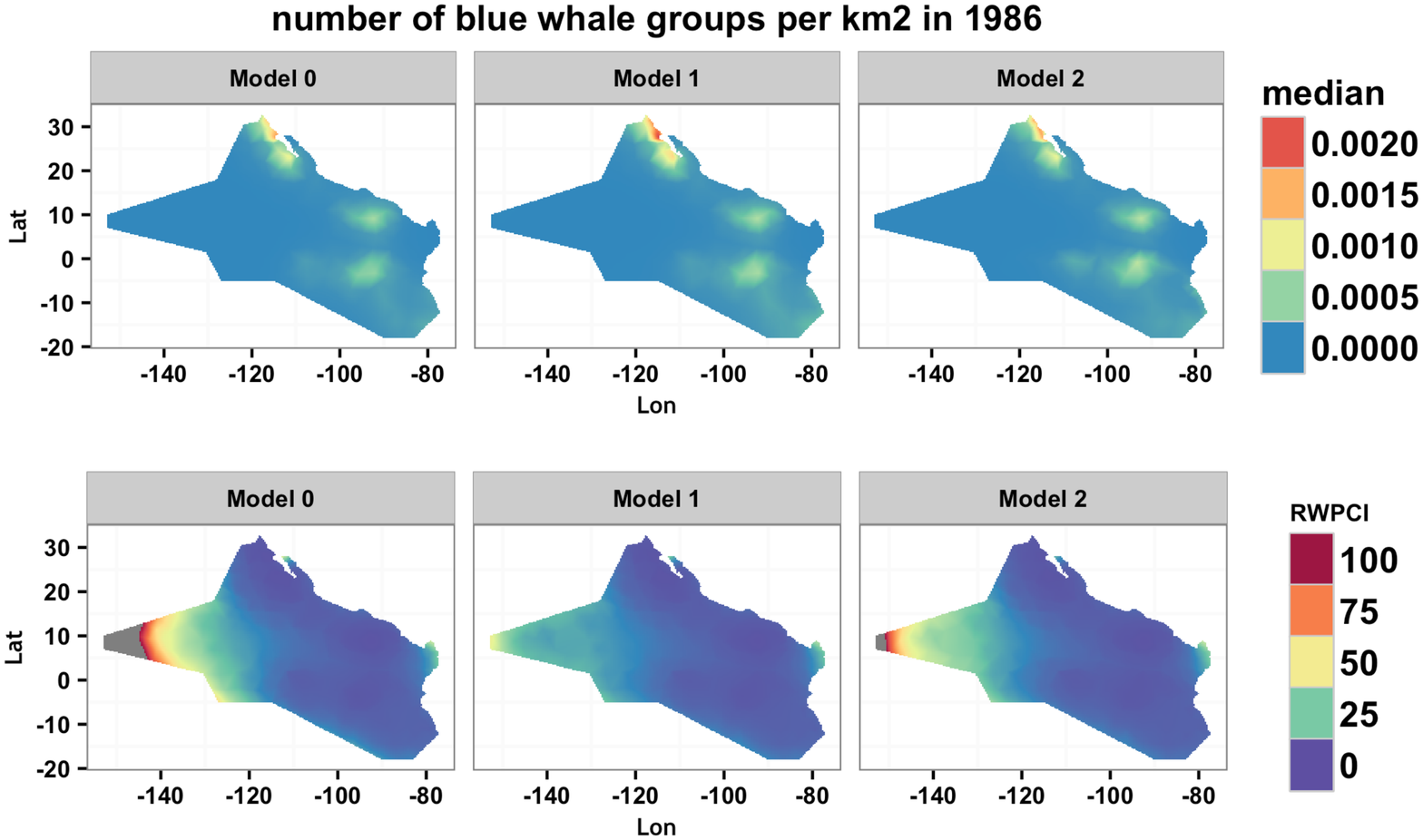}
	\vspace{-0.96cm}
	\caption{The posterior median (top) and RWPCI from~\eqref{eq:RWPCI} (bottom) of the ETP blue whale groups density in 1986 using Models 0, 1, and 2 in Table~\ref{tab:betasAllmodel}. The RWPCI colour palette is cut off at 100 to exclude the extreme values at the western corner of the ETP survey area. }
	\label{fig:fvmedian1986}
\end{figure}

The posterior median of blue whale density, $\lambda(\bm{s};t)$, for year 1986 is shown in the top panel of Fig~\ref{fig:fvmedian1986} for Models 0, 1 and 2, respectively. The top three plots of 
Fig~\ref{fig:fvmedian1986} are very similar, with areas of higher blue whale group density in the north off the coast of Baja California, in the area of the Costa Rica Dome off the coast of Central America, and in the south-east in the vicinity of the Galapagos Islands. This pattern of the blue whale group density is consistent across all the models implemented, and reflects what we observe in the sightings data in the right panel of Fig~\ref{fig:effortsight}. This observed spatial pattern is also in general agreement with previous analysis of blue whale sighting data in the ETP \citep{Forney+al:2012, Pardo+al:2015}. Similar plots for 1986--2007 (omitting the very strong El Ni\~{n}o year 1997), are given in Supplement~\ref{appx:densityests}. 

We use the relative width of the 95\% posterior credible interval (RWPCI) as a measure of the relative uncertainty for the predicted $\lambda(\bm{s};t)$ of the ETP survey area. We define the RWPCI as the inter-quartile range divided by the median,
\begin{align}
\mbox{RWPCI} &= (Q_3-Q_1)/Q_2.
\label{eq:RWPCI}
\end{align}
When the posterior distribution is approximately Gaussian, the RWPCI is about $1.35$ times the ratio of the posterior standard deviation to the posterior median. 
The bottom panel of Fig~\ref{fig:fvmedian1986} shows the spatial structure of the RWPCI in 1986 for each of the three models, and this pattern persists across years (see Supplement~\ref{appx:densityests}). The far west of the survey region has very high relative uncertainty because it is close to the edge of the mesh boundary shown in Fig~\ref{fig:effortsight} and there are no sightings in that area. The spatial random field has high uncertainty in this area: regions of low $\lambda(\bm{s};t)$ tend to have higher uncertainty associated with the latent field. The slowly varying standard deviation of the latent field in Fig~\ref{fig:LFallmodels} is likely due to a combination of large spatial range (see Fig~\ref{fig:Matern} of Supplement~\ref{appx:MaternDetfun}) and the fact that the observed point pattern is not very informative about the latent field.

\begin{figure}[h!]
	\centering
	\includegraphics[width=.65\linewidth]{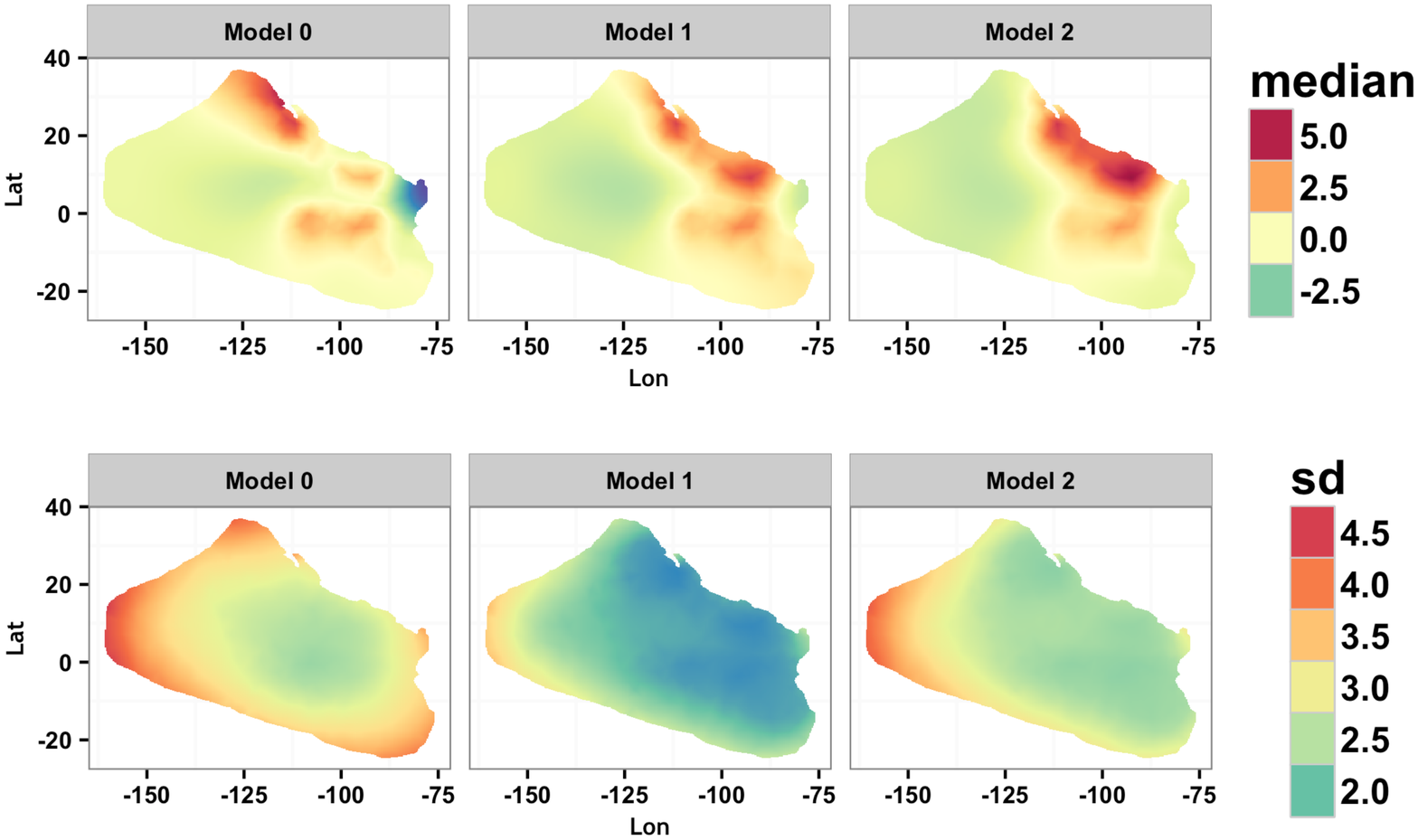}
	\caption{The posterior median and standard deviation of the latent field (\ref{eq:fem}) for Models 0, 1 and 2.}
	\label{fig:LFallmodels}
\end{figure}

\subsubsection{SPDE parameters and detection function}
Prior sensitivity tests of the SPDE parameters showed the posterior median of $\lambda(\bm{s};t)$ to be less sensitive to prior specification than is its variance. Details of the SPDE prior specification are given in Supplement~\ref{appx:spdeprior}. Fig~\ref{fig:SPDEhyperTwoModels} displays the posterior densities of the SPDE parameters for Models~0, 1 and 2 using the same prior. 

\begin{figure}
\centering
\begin{tabular}{cc}
\includegraphics[width=.25\linewidth]{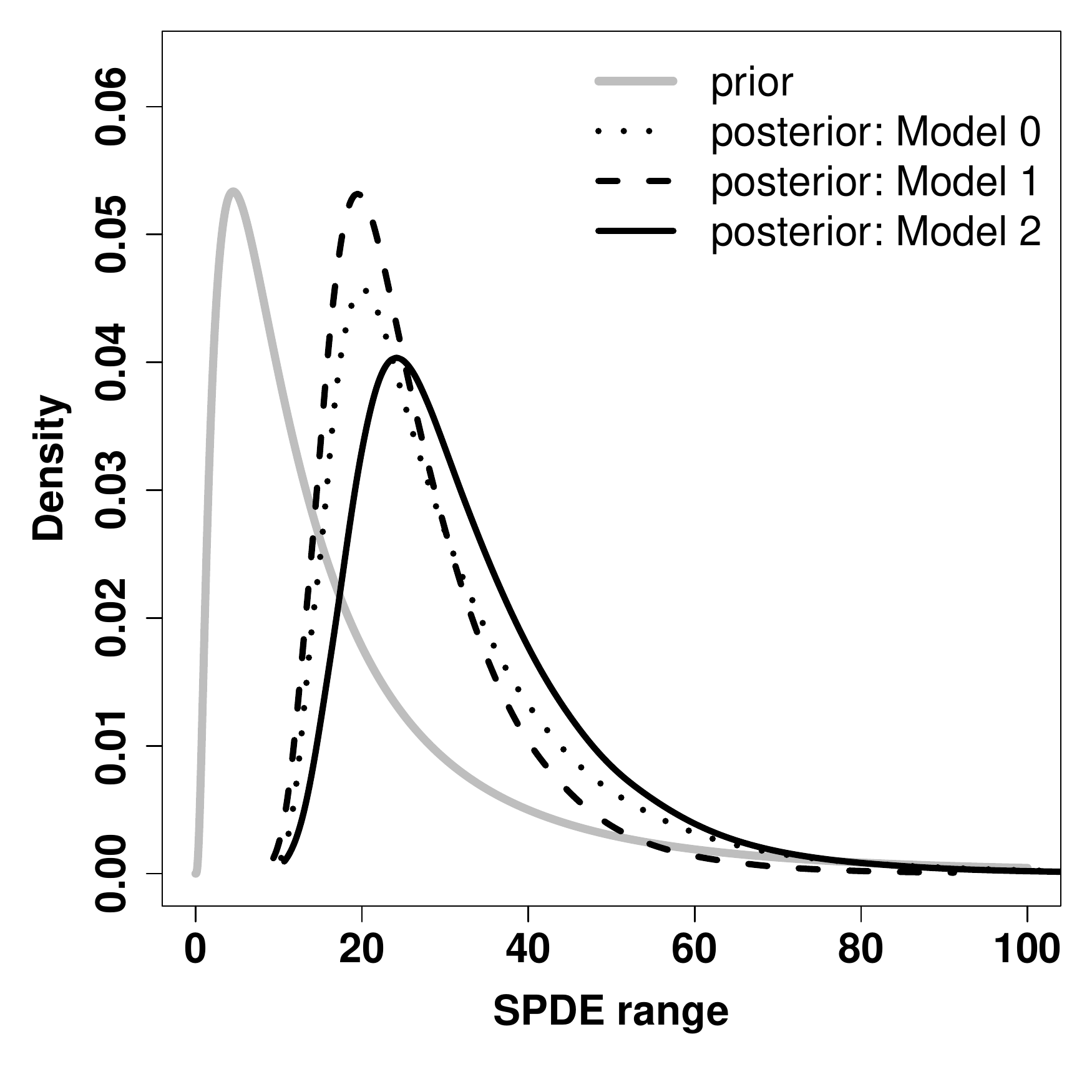}&
\includegraphics[width=.25\linewidth]{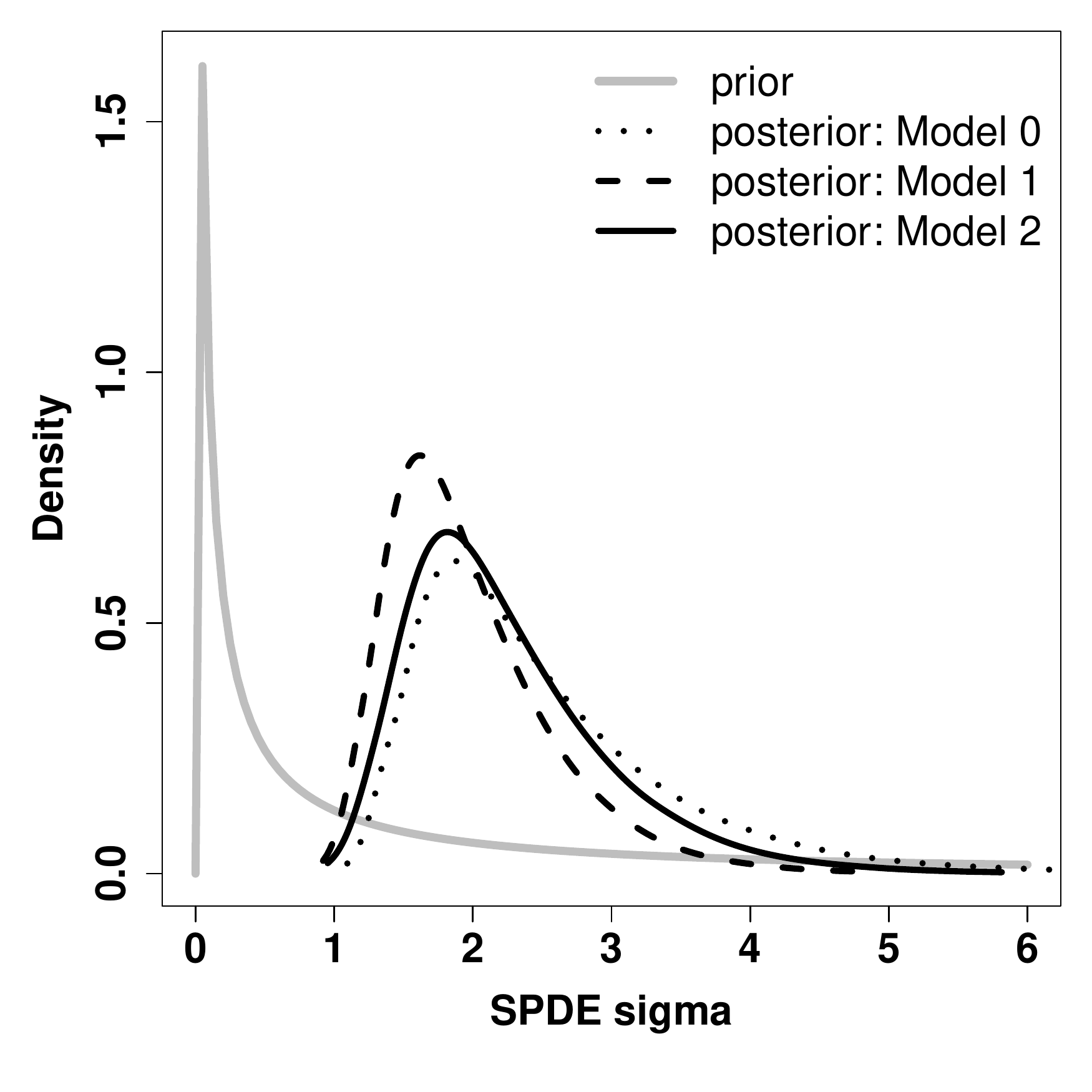}
\end{tabular}
\caption{The posterior densities of the SPDE parameters using Models 0, 1 and 2. The left panel is for the range parameter $\rho$ (see Section~\ref{sec:basicSPDE} for its definition), and the right panel for the marginal standard deviation $\sigma_{\xi}$ in~(\ref{eq:SPDE}). 
}
\label{fig:SPDEhyperTwoModels}
\end{figure}

The large range of the Mat{\'e}rn covariance function is consistent with the latent Gaussian random field ($\xi$ of~\eqref{eq:logLambda}) shown in Fig~\ref{fig:LFallmodels}. There is little difference among the models for either the posterior detection function or the 95\% credible band (see Fig~\ref{fig:postgx} of Supplement~\ref{appx:MaternDetfun}).

\subsection{Exploratory model checking}
\newcommand{\s}{{\bm{s}}}
Let $\eta(\s,t)$ denote the log-intensity
defined by the fixed effects and random field components of
\eqref{eq:logLambda},
\begin{align*}
  \eta(\s,t)&=\log[\lambda(\s;t)]=\bm{x}(\s,t)^\top\bm{\beta} +
  \xi(\s,t) .
\end{align*}
To investigate the role of the components and the possibility of
confounding, we consider the variability around the posterior mean of
the overall averages of $\eta(\s,t)$;
\begin{align*}
  M_{\eta}&=\frac{1}{|\Omega|\times |\mathbb{T}|} \int_{\s\in \Omega} \int_{t\in\mathbb{T}} E[\eta(\s,t) | \bm{Y}]\md t \md\s,
\end{align*}
and similarly for the components, so that $M_\eta=M_{\beta}+M_\xi$.
The posterior expected squared deviation of $\eta(\s,t)$ from
$M_\eta$ can be split into contributions from the fixed
effects $\bm{x}(\s,t)^\top\bm{\beta}$ and the random field
$\xi(\s,t)$;
\begin{align*}
  V_\eta(\s,t) = E\{[\eta(\s,t)-M_\eta]^2|\bm{Y}\} &=
  E\{[\bm{x}(\s,t)^\top\bm{\beta}-M_{\beta}]^2|\bm{Y}\}
  + E\{[\xi(\s,t)-M_\xi]^2|\bm{Y}\}
  \\&\phantom{=\,}
  +
  2 E\{[\bm{x}(\s,t)^\top\bm{\beta}-M_{\beta}] [\xi(\s,t)-M_\xi]|\bm{Y}\}
  \\&= V_\beta(\s,t) + V_\xi(\s,t) + 2C_{\beta,\xi}(\s,t) .
\end{align*}
For all models, $\xi(\s,t)$ is constant over time, and we define the averages across time, $V_\eta(\s,\mathbb{T})$, $V_\beta(\s,\mathbb{T})$, and $V_\xi(\s,\mathbb{T})$, shown in Fig~\ref{fig:Vmodel1} for Model 1 and in Supplement~\ref{appx:latenfixed} for all models. It is clear that the random field component $\xi(\s)$ captures information not available in the SST components.

The full space-time averages $V_\eta(\Omega,\mathbb{T})$,
$V_\beta(\Omega,\mathbb{T})$, and $V_\xi(\Omega,\mathbb{T})$ are the
variances when probing the posterior distributions at a uniformly
chosen random locations on $\Omega\times\mathbb{T}$.  The remainder
term $C_{\beta,\xi}(\Omega,\mathbb{T})$ is the posterior covariance
between the fixed effect and random field contributions to the
variability. We also define the corellation
$\rho_{\beta,\xi}(\Omega,\mathbb{T})=C_{\beta,\xi}(\Omega,\mathbb{T})/\sqrt{V_\beta(\Omega,\mathbb{T})V_\xi(\Omega,\mathbb{T})}$. A
large negative value for the covariance or correlation indicates
confounding. Table~\ref{tab:gofsummary} shows the space-time averages,
covariance and correlation for the three models. The correlations are
not very small, suggesting that there is some confounding, although it
is not severe.  While these diagnostics do not give direct guidance
for model selection, they highlight the clear contribution of the
random field component of each model.
\begin{table}[ht]
  \centering
  \caption{The posterior space-time averages $V_\eta(\Omega,\mathbb{T})$, $V_\beta(\Omega,\mathbb{T})$, $V_\xi(\Omega,\mathbb{T})$, covariance $C_{\beta,\xi}(\Omega,\mathbb{T})$ and correlation $\rho_{\beta,\xi}(\Omega,\mathbb{T})$ for Models 0, 1 and 2.}
  \label{tab:gofsummary}        	
  \centering
  \begin{tabular}{rrrrrr}
    \hline
    & $V_\eta(\Omega,\mathbb{T})$ &$V_\beta(\Omega,\mathbb{T})$ & $V_\xi(\Omega,\mathbb{T})$ & $C_{\beta,\xi}(\Omega,\mathbb{T})$ & $\rho_{\beta,\xi}(\Omega,\mathbb{T})$\\ 
    \hline
    Model 0 & 7.32 & 10.05 & 9.86 & -6.29 & -0.63 \\
    Model 1 & 5.04 &  3.19 & 5.97 & -2.06 & -0.47 \\
    Model 2 & 6.82 &  7.51 & 9.40 & -5.05 & -0.60 \\
    \hline
  \end{tabular}
\end{table}


\begin{figure}
\centering
\includegraphics[width=.7\textwidth]{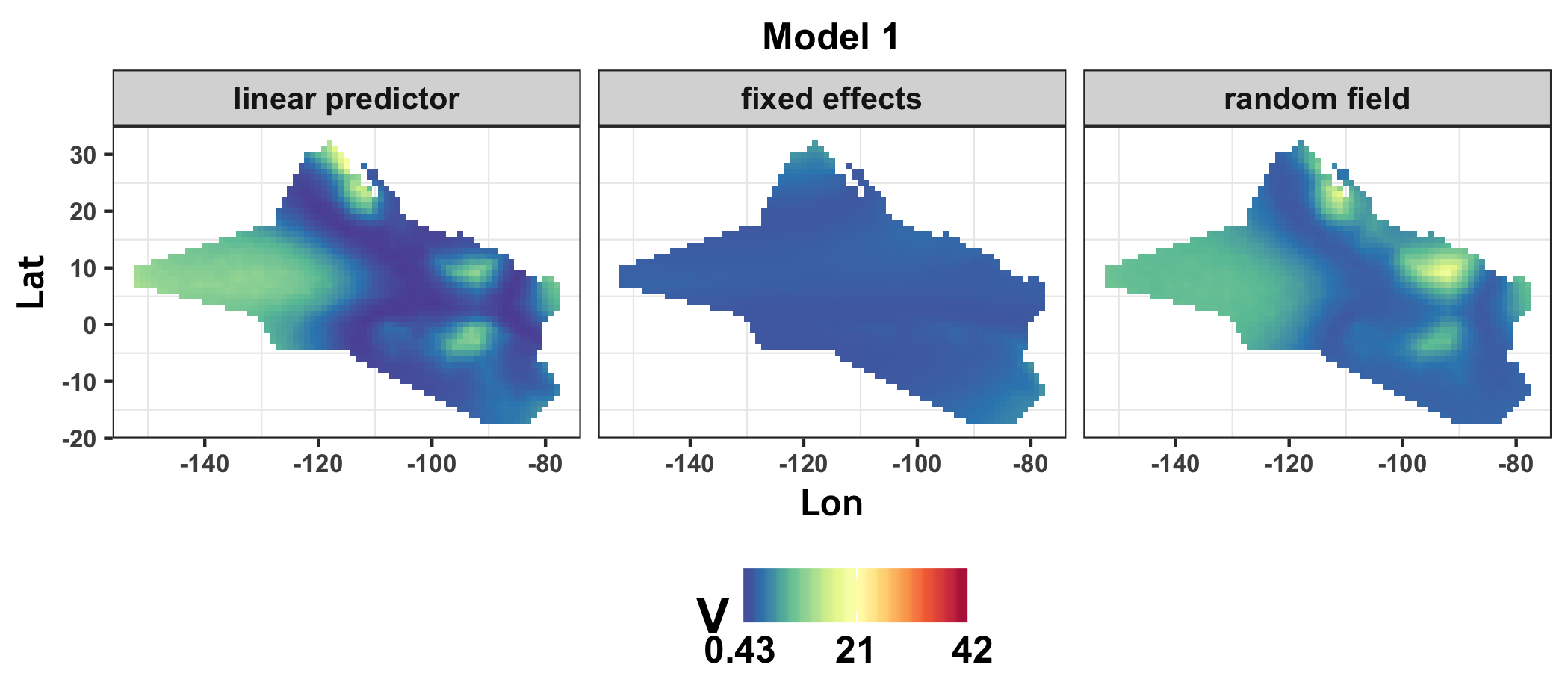}
\caption{Variability measures $V_\eta$, $V_\beta$, and $V_\xi$ for Model 1.}
\label{fig:Vmodel1}
\end{figure}

\section{Discussion}\label{sec:discuss}
Unlike previous methods used to analyse these and similar survey data, our spatio-temporal point process model preserves the sighting locations, models the effect of explanatory variables continuously in space, and models spatial correlation that cannot be explained by such variables. It generalizes the approach of  \citet{Johnson+al:2010}, which models density as a nonhomogeneous Poisson process, using actual sighting locations, but neglecting residual spatial correlation. Unlike \citet{Johnson+al:2010}, we model residual spatial intensity. It also generalises the approach of \citet{Pardo+al:2015}, who included a model for residual spatial intensity in their analysis of ETP blue whale data, but with no spatial structure on their residual model. We found substantial evidence for residual spatial structure in our analysis

It is rarely the case that spatial data are independent, and assuming independence when data are dependent can lead to biased variance estimation, spurious significance of covariates, and overfitting \citep{Cressie:1993, Hanks+al:2015}. Use of a GMRF allows us to model spatially autocorrelated random effects, and model patterns in residuals that cannot be explained by available covariates. As shown in Section~\ref{sec:ETPresults}, the spatial pattern captured by the GMRF in Fig~\ref{fig:LFallmodels} plays an important role in estimating the spatial distribution of ETP blue whale groups, shown in Fig~\ref{fig:fvmedian1986}. Because the underlying mechanisms that dictate the distribution of blue whales in space and time are probably quite complex, it is unlikely that SST alone could adequately explain the distribution, so that drawing inferences about the effect of SST based on a model without modeling spatial correlation may result in misleading biological interpretations.

The analysis of \citet{Pardo+al:2015} modeled blue whale density spatially as a function of absolute dynamic topography (ADT), which, like SST, predicted fewer blue whales in warmer regions. Because the model did not separate the temporal and spatial effects of ADT, large changes in ETP blue whale abundance were predicted from year to year, with few whales in warm (El Ni\~{n}o) years and many whales in cool (La Ni\~{n}a) years.  Because blue whales have long life spans and reproduce slowly, and because tagging has shown that blue whales migrate to tropical waters every year, regardless of El Ni\~{n}o variations \citep[see tracks in][for example]{Bailey+al:2009}, high interannual variation in true abundance seems unlikely. The hierarchical centering scheme in Section~\ref{sec:centeringSST} separates the temporal and spatial effects of SST and accommodates situations in which whales make choices about habitat use \textit{relative} to the other choices available to them, and this leads to what is arguably a more biologically plausible model with less interannual variation. While our estimates are consistent with true blue whale group density being unchanged over the surveys, point estimates of this density do tend to be higher in warmer years. This is unexpected and warrants further investigation. \citet{Pardo+al:2015} argue that ADT is a better predictor of blue whale density than SST because ADT contains information about subsurface as well as surface conditions. Notwithstanding this, our model is able to pick up structure in the data beyond that which can be attributed to SST. For example, \citet{Pardo+al:2015} predict high denities on the Costa Rica Dome (approximately 10$^o$N, 90$^o$E) on the basis of ADT; we do the same by means of the GMRF (see Fig~\ref{fig:LFallmodels}) even though SST does not suggest high densities here. By also modeling spatial autocorrelation, our model does not run the risk of drawing biased inference about the effects of explanatory variables (SST here) due to unmodeled correlation. We found that the estimated Gaussian random field is somewhat correlated with the fixed effects assocaited with SST. As a result, the interpretation of the fixed effects is not as clearcut as it would be were the Gaussian random field and fixed effects independent. 

Considering our models in the more general context of point process modelling, the data structure we consider here differs from the point patterns typically analyzed in the point process literature \citep[but see][]{Waagepetersen+Schweder:2006}. These usually comprise a point pattern that has been observed completely in a finite observation window that is a subset of $\R^2$, say. Unless finite point processes are explicitly considered the standard assumption is that the point process continues in the same way outside the observation window. For interpretation, this implies that the analysis is only informative if the processes of interest are operating at a spatial scale that is captured within the (frequently single) subsample that is available. Further, there is an additional assumption that every point in the observation window has been observed, so that the detection probability is one within the observation window and zero elsewhere.

Our method extends such methods to deal with situations in which the processes of interest reflected in a spatial pattern, such as habitat preference, operates at a larger spatial scale than the sampled regions, when it may be impossible to fully sample an area that captures that scale. It also accommodates situations in which detection probability is unknown and not one, even within the sampled region. In wildlife sampling literature this has often been dealt with in two stages, first estimating detection probability and then estimating spatial distribution conditional on the estimated detection probability. Our approach integrates the two, estimating detection probability simultaneously with the point process parameters.

We expect to see advances in spatio-temporal inference when there are covariates that affect both the thinning process and the density surface \citep{Dorazio:2012}. We also expect further development of methods to assess goodness of fit, as such methods are somewhat lacking for spatial and spatio-temporal inference. 

The point process model in Section~\ref{sec:modelsoverall} can also be extended to a marked point process model to incorporate group size in the model and allow detection probability to depend on group size. We also anticipate that our approach will be extended to deal with more complex observation processes and for other survey types -- for spatial capture-recapture sampling \citep{Borchers+Efford:08, Royle+Young:08} for example, for situations in which detection probabilities change over time, or when there is unknown spatially varying sampling effort.

\section*{Acknowledgements}
This research was funded by EPSRC grants EP/K041061/1 and EP/K041053/1. We thank the captains, crews and observers on the NOAA research vessels, and the support staff at the Southwest Fisheries Science Center, for the collection of line-transect data in the ETP over many years.

\bibliographystyle{apalike}
\bibliography{main}

\clearpage
\newpage

\appendix

\renewcommand\thefigure{\thesection.\arabic{figure}}    
\setcounter{figure}{0}    

\clearpage
\section{Some assumptions}\label{appx:LTassumptions}
\setcounter{figure}{0}
The assumptions referred to in Section~\ref{sec:linetransectlikelihood} are as follows:
\begin{enumerate}
  \item A team of observers is considered as a joint
    \emph{black box} system, and the aggregated detection properties
    are modeled.
\item Individual objects (animals or animal groups) are not uniquely identified, only their locations are observed.
\item For each segment, the observable regions behind the
  starting point and ahead of the endpoint are small compared with the length of the segment as a whole, and the partial overlap of segments at changes
  in path direction is negligible.
\item The time between any other segment overlap is large enough that
  the time-slice point patterns in the overlap region can be
  considered independent; the object curves are considered to be in
  equilibrium, and at least locally mixing faster than the time
  between revisits by the observer.
\end{enumerate}

\clearpage
\section{Some details of SPDE models}\label{appx:spde}
\setcounter{figure}{0}
As noted in the body of the paper, the results from \citet{Lindgren+al:2011} show how to take advantage of the connection between Gaussian Markov random fields of
graphs and stochastic partial differential equations in continuous
space.  The most basic such model is based on the following stochastic
partial differential equation (SPDE) defined on a 2-dimensional
spatial domain
\begin{align}
\label{eq:SPDE}
(\kappa^2 -  \nabla\cdot\nabla)[\tau \xi(\bm{s})] &= \mathcal{W}(\bm{s}),
\quad \bm{s} \in \R^2,
\end{align}
where $\nabla\cdot\nabla$ is the Laplacian, $\mathcal{W}(\bm{s})$ is
Gaussian spatial white noise, and $\tau,\kappa>0$ are variance and
range scaling parameters.  \citet{Whittle:1954, Whittle:1963} proved
that stationary solutions to (\ref{eq:SPDE}) are Gaussian random
fields (GRF) with Mat{\'e}rn covariance function,
\begin{align}
\label{eq:materncov}
\cov \left[\xi(\bm{s}), \xi(\bm{s}')\right]
& =
\sigma_{\xi}^2\, \kappa \, \|\bm{s}'-\bm{s}\|
\,{K}_1\left(\kappa \, \|\bm{s}'-\bm{s}\|\right),
 \quad \bm{s},\bm{s}'\in\R^2,
\end{align}
where $\sigma_\xi^2={1}/({4\pi\kappa^2\tau^2})$ is the marginal
variance, and $\mathcal{K}_1$ is the modified Bessel function of the
second kind and order 1. The corresponding correlation function is
\begin{align}
\label{eq:materncorr}
\mbox{cor} \left[\xi(\bm{s}), \xi(\bm{s}')\right]
& =
\kappa \, \|\bm{s}'-\bm{s}\|
\,{K}_1\left(\kappa \, \|\bm{s}'-\bm{s}\|\right),
 \quad \bm{s},\bm{s}'\in\R^2.
\end{align}
A measure of the spatial range can be
obtained from $\rho = {\sqrt{8}}/{\kappa}$, which is the distance
where the spatial correlation is approximately $0.13$. More complex models can be obtained by changing the operator order or allowing the parameters to depend on the location.  Spatio-temporal models can be constructed by either using a temporally continuous differential operator such as in the heat equation, or with auto-regressive constructions in discrete time, such as
\begin{align}
\label{eq:xist}
\xi(\bm{s}, t)=a\,\xi(\bm{s}, t-1) + {\omega(\bm{s}, t)},
\end{align}
where $|a|<1$ controls the temporal autocorrelation, and $
\omega(\bm{s}, t)$ are solutions to \eqref{eq:SPDE}, independent for
each $t$.  These constructions can also be directly applied to
non-Euclidean domains such as the sphere, making construction of
globally consistent random field models straightforward.

\clearpage
\section{Integration scheme linearisation}\label{appx:integration_projection}
\setcounter{figure}{0}
Consider a spatial integration scheme for a fixed time point.  Reorganise
the integration points from Section~\ref{sec:numericalintegration} so
that $\mv{u}_{kj}$ is integration point number $j$ falling in
triangle $k$, with $k=1,\dots,K$ and $j=1,\dots,J_k$. The
corresponding integration weights are $w_{kj}$.  For a given triangle
we then obtain a linear approximation of function evaluations,
\begin{align*}
f(\bm{u}_{kj}) \approx \sum_{i=1,2,3} b_{kji}f_{ki},
\end{align*}
where the $b_{kji}$ are the Barycentric coordinates~\citep{Farin:2002} of $(\bm{u}_{kj})$ with respect to the triangle and the $f_{ki}$ denote the function $f$ evaluated at the triangle vertices. It follows that the sum approximating the integration over a fixed triangle $k$ can be carried out by three function evaluations,
\begin{align*}
\sum_{j=1}^{J_k} w_{kj} f(\bm{u}_{kj}) &\approx \sum_{j=1}^{J_k} w_{kj} \sum_{i=1,2,3} b_{kji}f_{ki} \\
&= \sum_{i=1,2,3} \left[\sum_{j=1}^{J_k} w_{kj}  b_{kji} \right] f_{ki}  \\
&= \sum_{i=1,2,3} \wt{w}_{ki} f_{ki}
\end{align*}
with weights $\wt{w}_{ki} = \sum_{j=1}^{J_k} w_{kj} b_{kji}$. Furthermore, vertices are shared among triangles. That is, $f_{a\cdot}$ and $f_{b\cdot}$ of two triangles $a$ and $b$ might refer to evaluations at the same mesh vertex. Hence, we can simplify the sum over all triangles as follows. Index the mesh vertices by $v=1,\dots V$ and let $v \sim (k,i)$ denote that $v$ is the $i$-th vertex of triangle $k$. The (possibly empty) set of weights associated with vertex $v$ then becomes $W_v=\{\wt{w}_{ki};\,v\sim (k,i)\}$ and we can write the full integral approximation as
\begin{align*}
\sum_{k=1}^K \sum_{j=1}^{J_k} w_{kj} f(\bm{u}_{kj}) &\approx \sum_{v=1}^V \ol{w}_vf_v,
\end{align*}
where $f_v$ denotes a function evaluation at vertex $v$ and $\ol{w}_v = \sum_{w \in W_v} w$.

\clearpage
\section{Sea surface temperature data}\label{appx:plotSST}
\setcounter{figure}{0}

\begin{figure}[h!]
\centering
\includegraphics[width=0.96\textwidth]{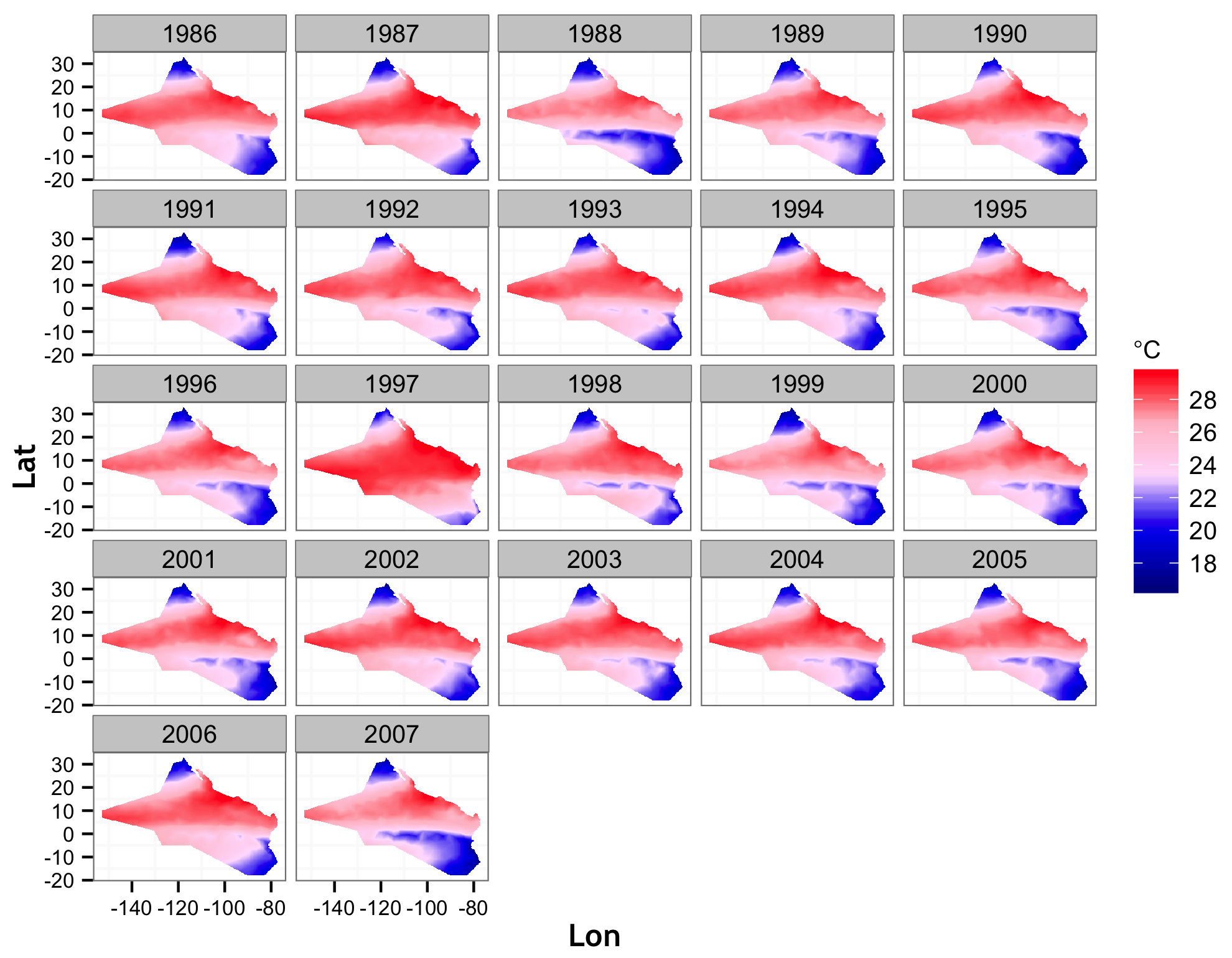}
\caption{The SST data before centering. For each location and year (1986--2007), the SST is averaged over July to December.}
\label{fig:SSTorig}
\end{figure}

In this supplement, we give details of the sea surface temperature (SST) in the form of plots. The uncentered SST data are shown in Figure~\ref{fig:SSTorig}. 

\begin{figure}
\centering
\includegraphics[width=0.9\textwidth]{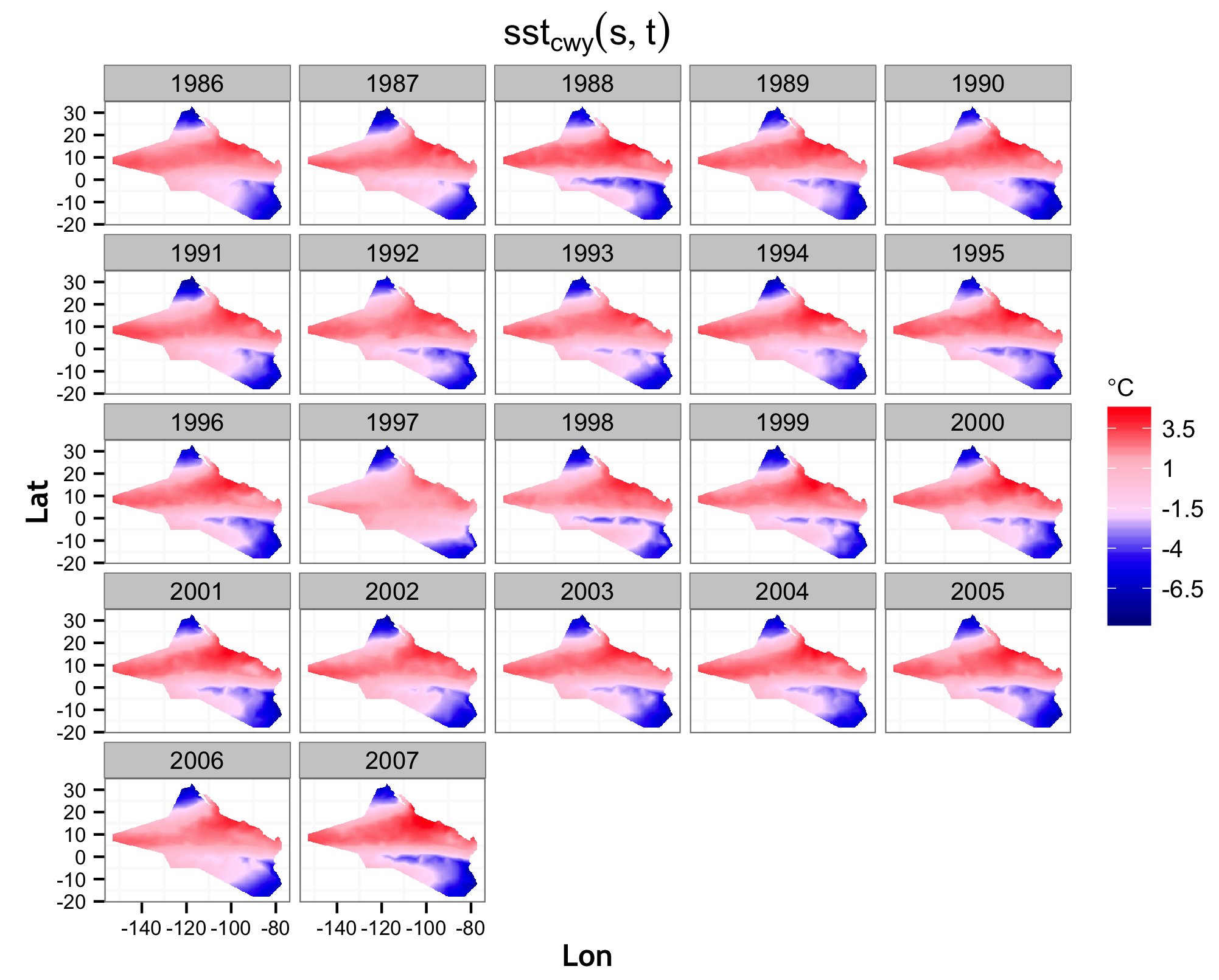}
\caption{The SST centered within year for the ETP survey area when using the within-year centering scheme. ${\sst}_{cwy}(\bm{s},t)$ is given by \eqref{eq:SSTcwy}}
\label{fig:SSTcwy}
\end{figure}

\begin{figure}
\centering
\includegraphics[width=0.56\textwidth]{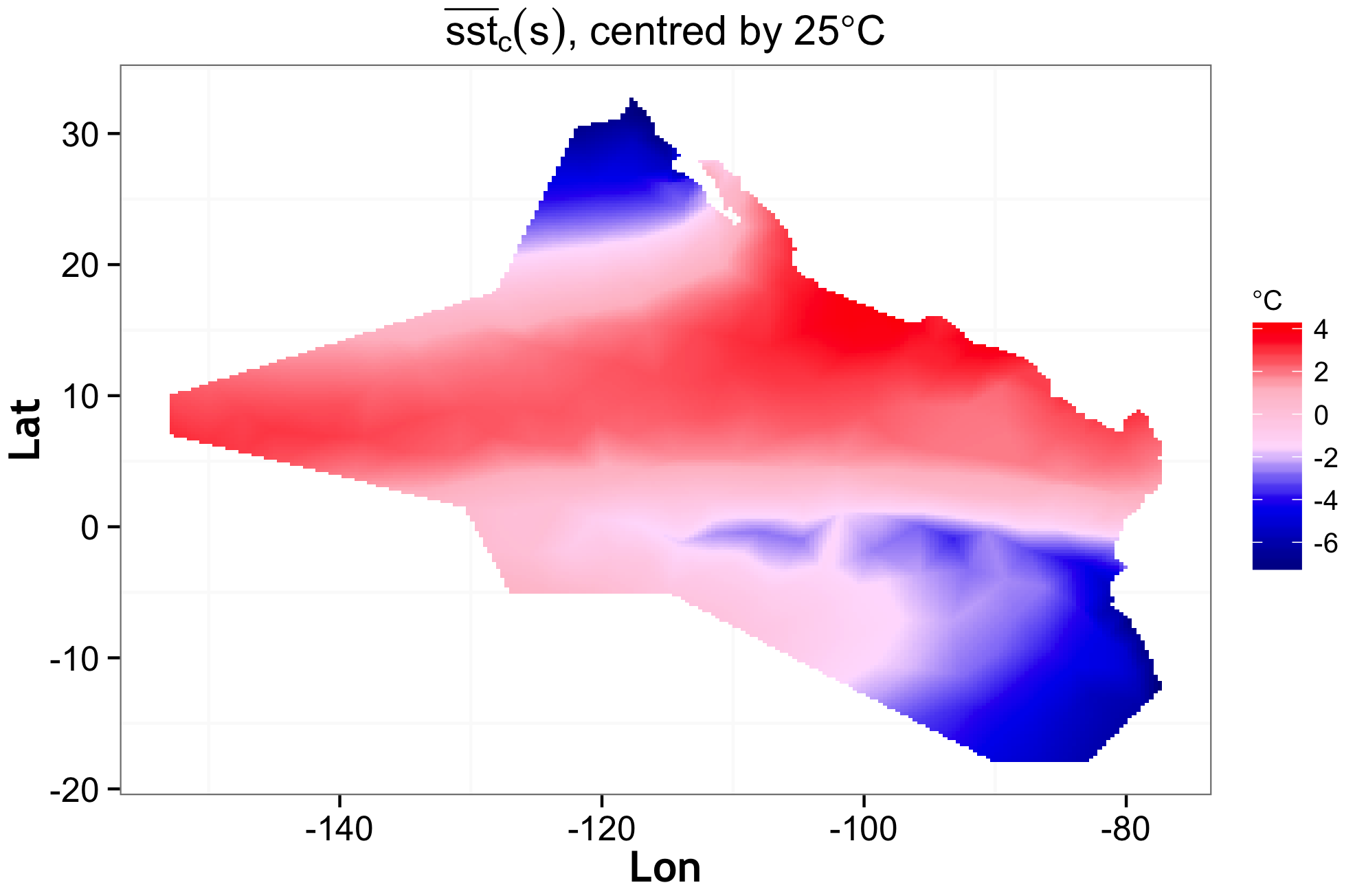}
\caption{The time invariant spatial SST pattern $\overline{\sst}_c(\bm{s})$ given by \eqref{eq:SSTcLoc}, where the overall average SST is $\overline{\sst} = \int_{\Omega\times\mathbb{T}}\sst(\bm{s}, t) \md\bm{s} \md t \big/(|\Omega|\times |\mathbb{T}|) \approx 25 \mbox{ \degree C}$.}
\label{fig:barSSTcLoc}
\end{figure}

\begin{figure}
\centering
\includegraphics[width=0.9\textwidth]{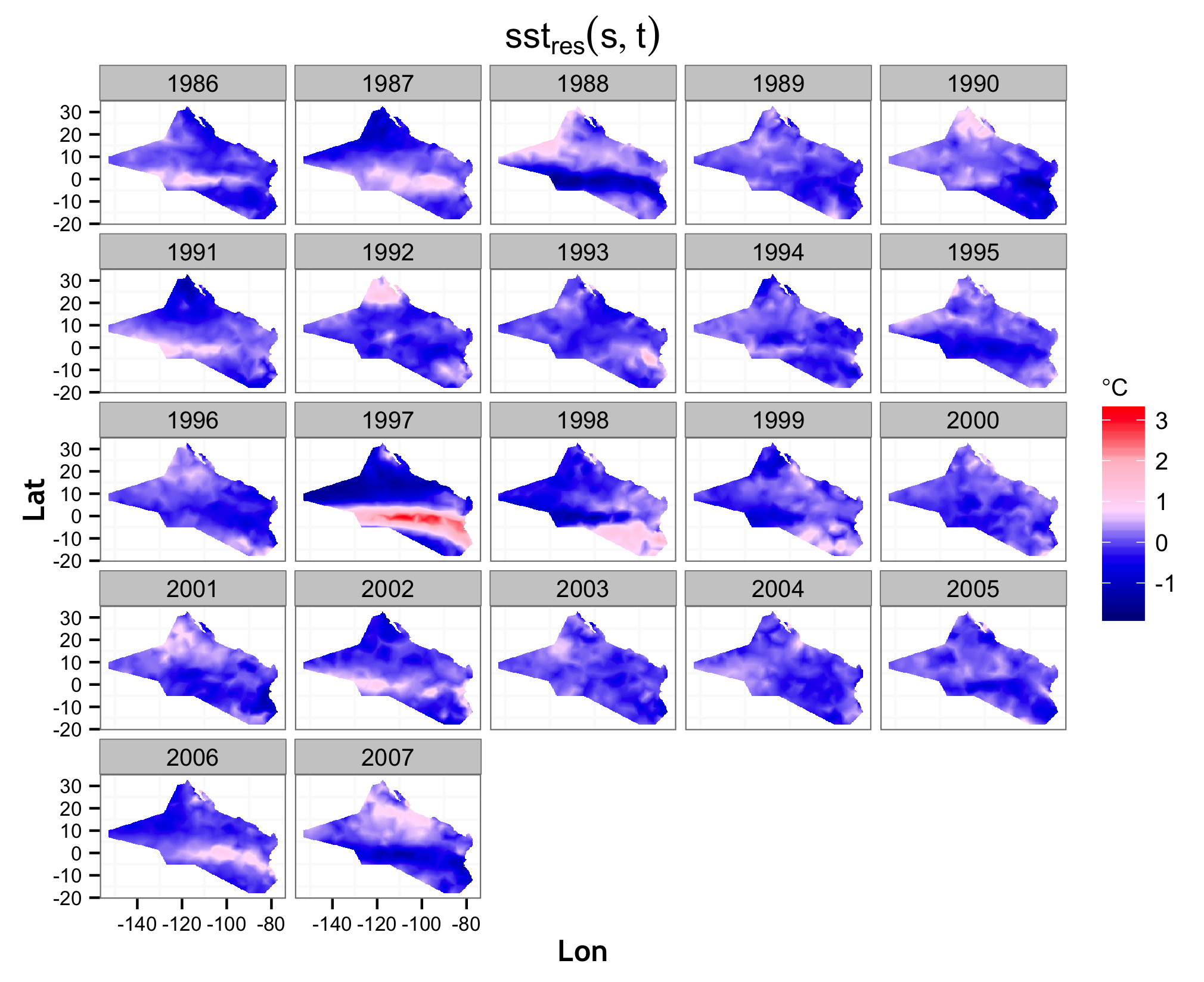}
\caption{The SST residuals when using the space-time centering scheme. $\sst_{res}(\bm{s}, t)$ is given by (\ref{eq:SSTres}).}
\label{fig:SSTres}
\end{figure}

SST is centered using the within-year and space-time centering schemes described in Section~\ref{sec:centeringSST}. The two components of SST when using the within-year centering scheme are the $\overline{\sst}_{c}(t)$ in Fig~\ref{fig:barSSTcYear} and ${\sst}_{cwy}(\bm{s},t)$ in Fig~\ref{fig:SSTcwy}, which have been defined in \eqref{eq:SSTcYr} by \eqref{eq:SSTcwy}, respectively. For the space-time centering scheme, in addition to $\overline{\sst}_{c}(t)$ in Fig~\ref{fig:barSSTcYear}, the remaining two components are displayed in Fig~\ref{fig:barSSTcLoc} and~\ref{fig:SSTres}, where Fig~\ref{fig:barSSTcLoc} shows the spatial pattern of SST averaged over all years, and Fig~\ref{fig:SSTres} shows the residual SST. After subtracting the SST averaged over space \eqref{eq:SSTcYr} and SST averaged over time \eqref{eq:SSTcLoc} from the SST displayed in Fig~\ref{fig:SSTorig}, the residual SST in Fig~\ref{fig:SSTres} shows residual patterns related to El Ni\~{n}o/La Ni\~{n}a oscillations, most clearly with the strong El Ni\~{n}o of 1997.

\clearpage
\section{Plots of estimated blue whale density}\label{appx:densityests}
\setcounter{figure}{0}

\begin{figure}[h!]
	\centering
	\includegraphics[width=.8\linewidth]{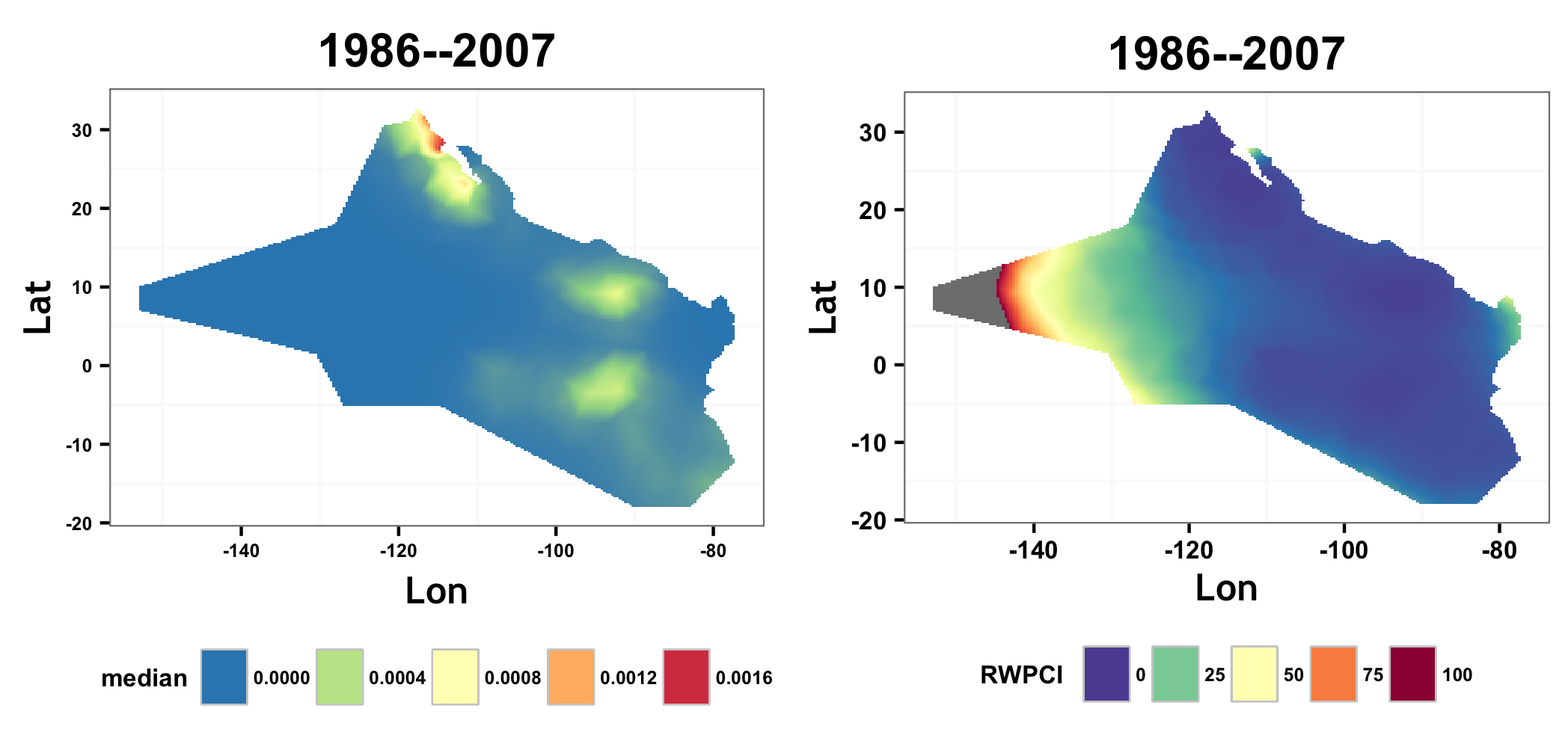}
	\caption{Posterior density of the blue whale groups using Model 0 in Table~\ref{tab:betasAllmodel}. The left panel displays the posterior median of the number of blue whale groups per square kilometre, and the right panel displays the relative width of the posterior credible interval given by (\ref{eq:RWPCI}). Model 0 does not incorporate any temporal information, these estimate apply to all years.}
	\label{fig:Model0PMandPCI}
\end{figure}

\begin{figure}
	\centering
	\includegraphics[width=1\linewidth]{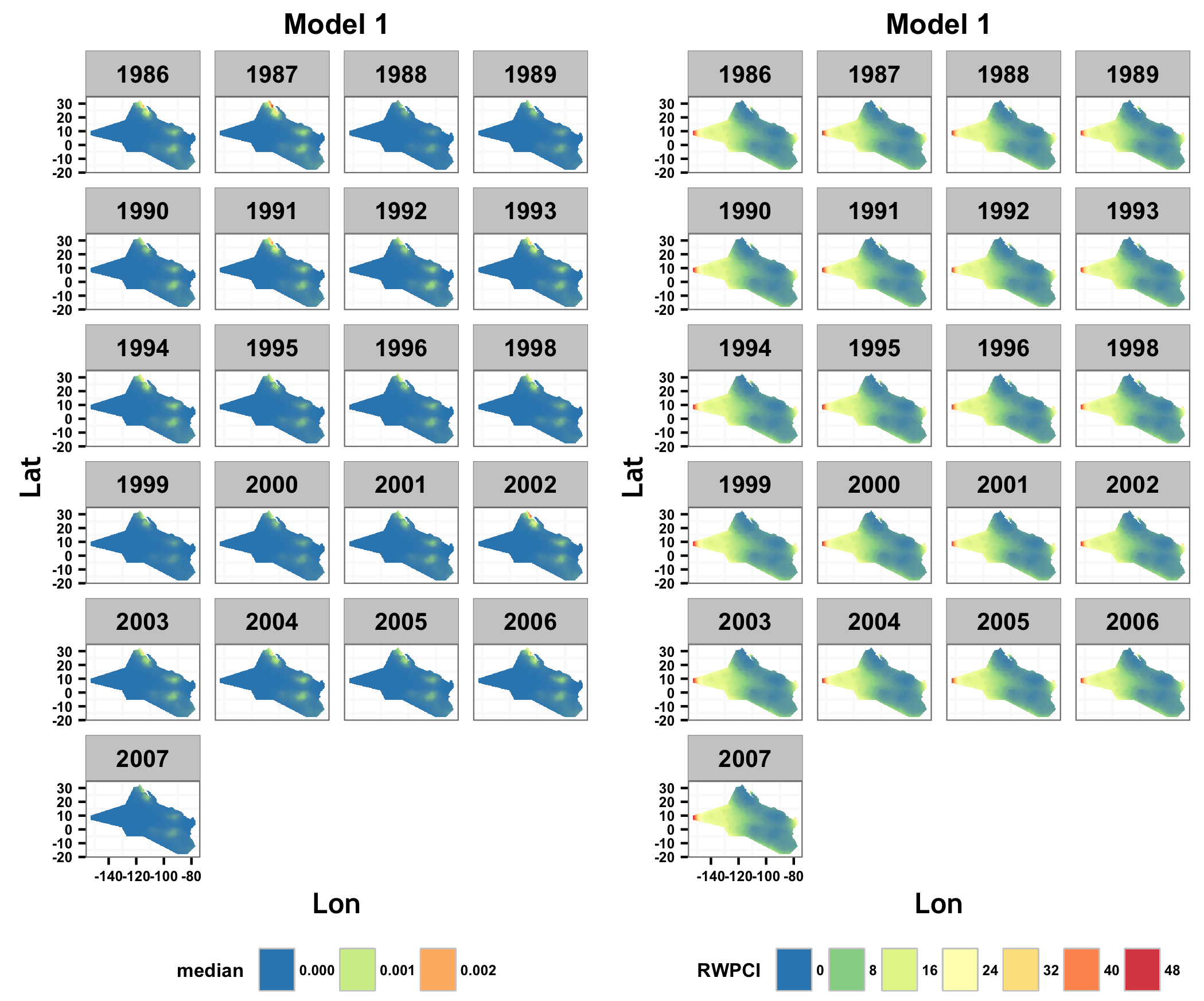}
	\caption{Posterior density of the blue whale groups using Model 1 in Table~\ref{tab:betasAllmodel}, years 1986--2007 (except 1997): the left panel displays the posterior median of the number of blue whale groups per square kilometre, and the right panel displays the relative width of the posterior credible interval given by (\ref{eq:RWPCI}).}
	\label{fig:Model1.1PMandPCI}
\end{figure}

\begin{figure}
	\centering
	\includegraphics[width=1\linewidth]{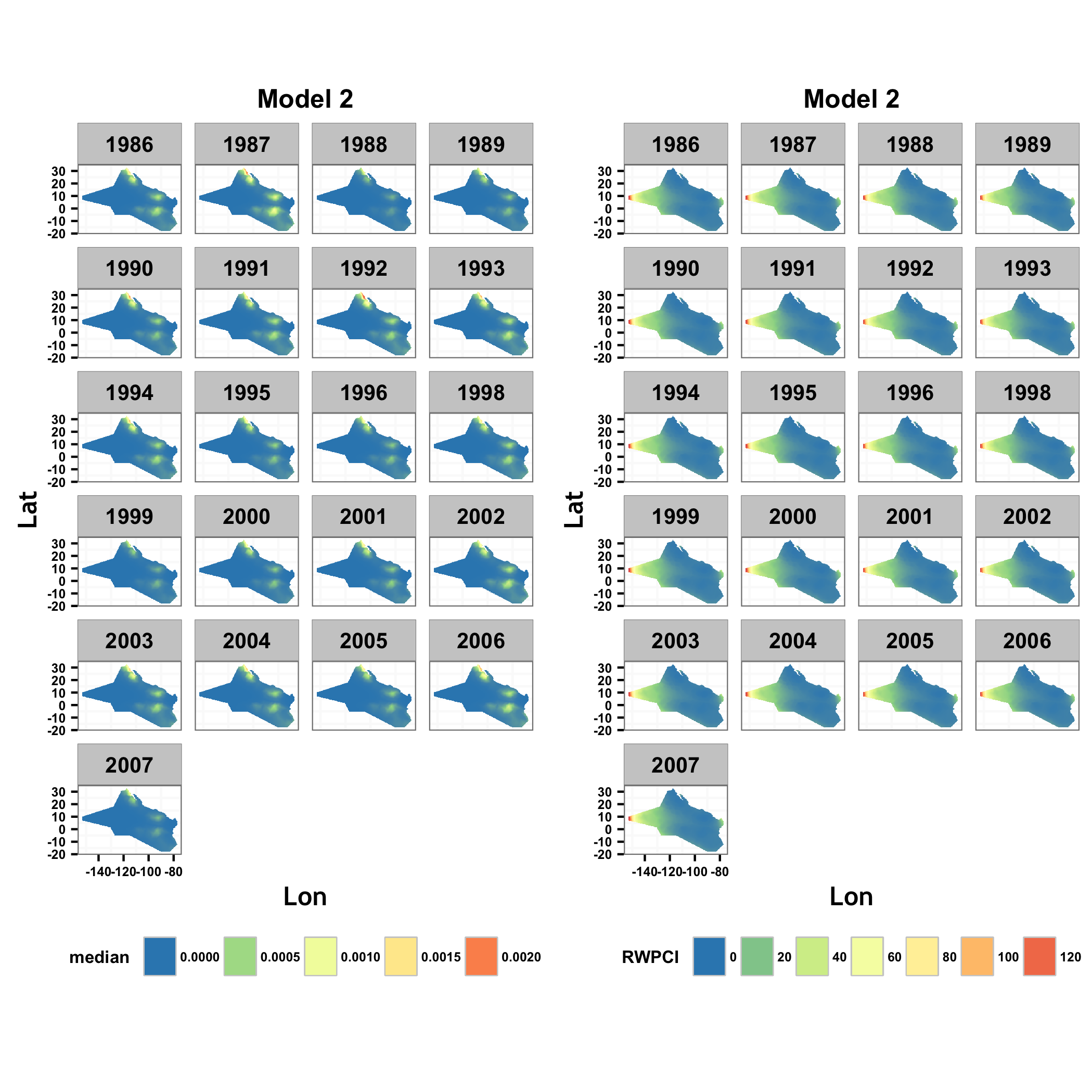}
	\caption{Posterior density of the blue whale groups using Model 2 in Table~\ref{tab:betasAllmodel}, years 1986--2007 (except 1997): the left panel displays the posterior median of the number of blue whale groups per square kilometre, and the right panel displays the relative width of the posterior credible interval given by (\ref{eq:RWPCI}). }
	\label{fig:Model2.1PMandPCI}
\end{figure}

Plots of the posterior medial blue whale group density are shown in Figures~\ref{fig:Model0PMandPCI}, \ref{fig:Model1.1PMandPCI} and \ref{fig:Model2.1PMandPCI}, for Model 0, Model 1 and Model 1, respectively. In the case of Model 0 a single plot is shown for all years as this model has no temporal component. Posterior denstiy estimates for Models 1 and 2 vary by year because of their dependence on SST.

\clearpage
\section{SPDE parameters}\label{appx:spdeprior}
\setcounter{figure}{0}
\texttt{R-INLA} specifies the SPDE model using two internal parameters: $\kappa$ and $\tau$, both of which have an influence on the marginal variance of the random field $\sigma^2$,
\begin{align*}
\sigma^2 &= \frac{\Gamma(\nu)}{\Gamma(\alpha)(4\pi)^{d/2}\kappa^{2\nu}\tau^2},
\end{align*}
where $\nu$ is the smooth parameter and $\alpha =\nu  - d/2$  with $d$ being the dimension of the domain.
In \texttt{R-INLA}, the default value is $\alpha =2$. \citet{Whittle:1954} argues that it is more natural to use $\alpha=2$ for $d=2$ models than the fractional $\alpha=3/2$, which generates exponential covariances. \citet{Rozanov:1982} shows that using integers for $\alpha$ gives continuous domain Markov fields, for which the discrete basis representation in Section~\ref{sec:SPDEcomputation} is the easy to construct.  $\kappa>0$ itself is related to the range parameter, denoted by $\rho$ and $\rho=\sqrt{8\nu}/\kappa$. The range parameter $\rho$ is commonly defined as the distance at which the spatial correlation function falls close to zero for all $\nu>1/2$. Therefore, in our case of $d=2$, we use $\alpha=2$ and it follows that
\begin{align}
\label{eq:sigmarho_d2}
\sigma^2& = \frac{1}{4\pi \kappa^2\tau^2}\quad \mbox{and} \quad \rho=\frac{\sqrt{8}}{\kappa}.
\end{align}

Working directly with the SPDE parameters $\kappa$ and $\tau$ can be
difficult, as they both affect the variance of the field. It is often
more natural and interpretable to consider the standard deviation
$\sigma$ and spatial range $\rho$.
The aim is to construct a joint prior for the internal model parameters $\kappa$ and $\tau$ such that $\sigma$ and $\rho$ get independent log-Normal priors,
\begin{align*}
  \begin{cases}
\sigma \sim \pLN\left(\log\sigma_0, \,\sigma_{\sigma}^2\right), &\\
\rho\sim \pLN\left( \log\rho_0,\,\sigma_{\rho}^2\right) , &
 \end{cases}
\end{align*}
where $\sigma_0$ and $\rho_0$ are the prior medians.

First, we choose $\sigma_0$ and $\rho_0$ based on the problem
domain. From (\ref{eq:sigmarho_d2}), we obtain the corresponding
values for the internal parameters $\kappa$ and $\tau$ from $\kappa_0
= \sqrt{8}/{\rho_0}$ and $\tau_0 ={1}\big/\sqrt{4\pi \kappa_0^2 \sigma_0^2}$.
Then, if $\tau$ and $\kappa$ are parameterised through log-linear combinations of two parameters $\theta_1$ and $\theta_2$,
\begin{align}
\label{eq:internalTauKappa}
\begin{cases}
\log (\tau) = \log (\tau_0)  -\theta_1  + \theta_2, & \\
\log (\kappa)  = \log (\kappa_0) - \theta_2, &
\end{cases}
\end{align}
the $\sigma$ and $\rho$ parameters are related to $\theta_1$ and $\theta_2$ as
\begin{equation}
\label{eq:logtaurhoInternal_d2}
  \begin{cases}
\log(\sigma)=-\log(\sqrt{4\pi}\,\tau_0\kappa_0)+\theta_1, &\\
\log(\rho)=\log(\sqrt{8}/\kappa_0) + \theta_2. &
\end{cases}
\end{equation}
Thus, $\theta_1$ and $\theta_2$ separately control the standard
deviation and spatial range, respectively. Assigning independent
Normal distributions to $\theta_1$ and $\theta_2$ leads to the desired result, since $\log\sigma_0=-\log(\sqrt{4\pi}\,\tau_0\kappa_0)$ and $\log\rho_0=\log(\sqrt{8}/\kappa_0)$.

General guidance for how to choose the prior medians and variances is difficult to provide.  Here, we use $\sigma_0=1$ and $\sigma_\sigma^2=10$ for the standard deviation, and $\rho_0=\textrm{domainsize}/5$ and $\sigma_\rho^2=1$.  The domain size for the ETP study is$~62$, and a fifth is a reasonable portion of that.

\clearpage
\section{Correlation function and detection function posteriors}\label{appx:MaternDetfun}

\begin{figure}[hb]
\centering
\begin{tabular}{ccc}
\includegraphics[scale=0.2]{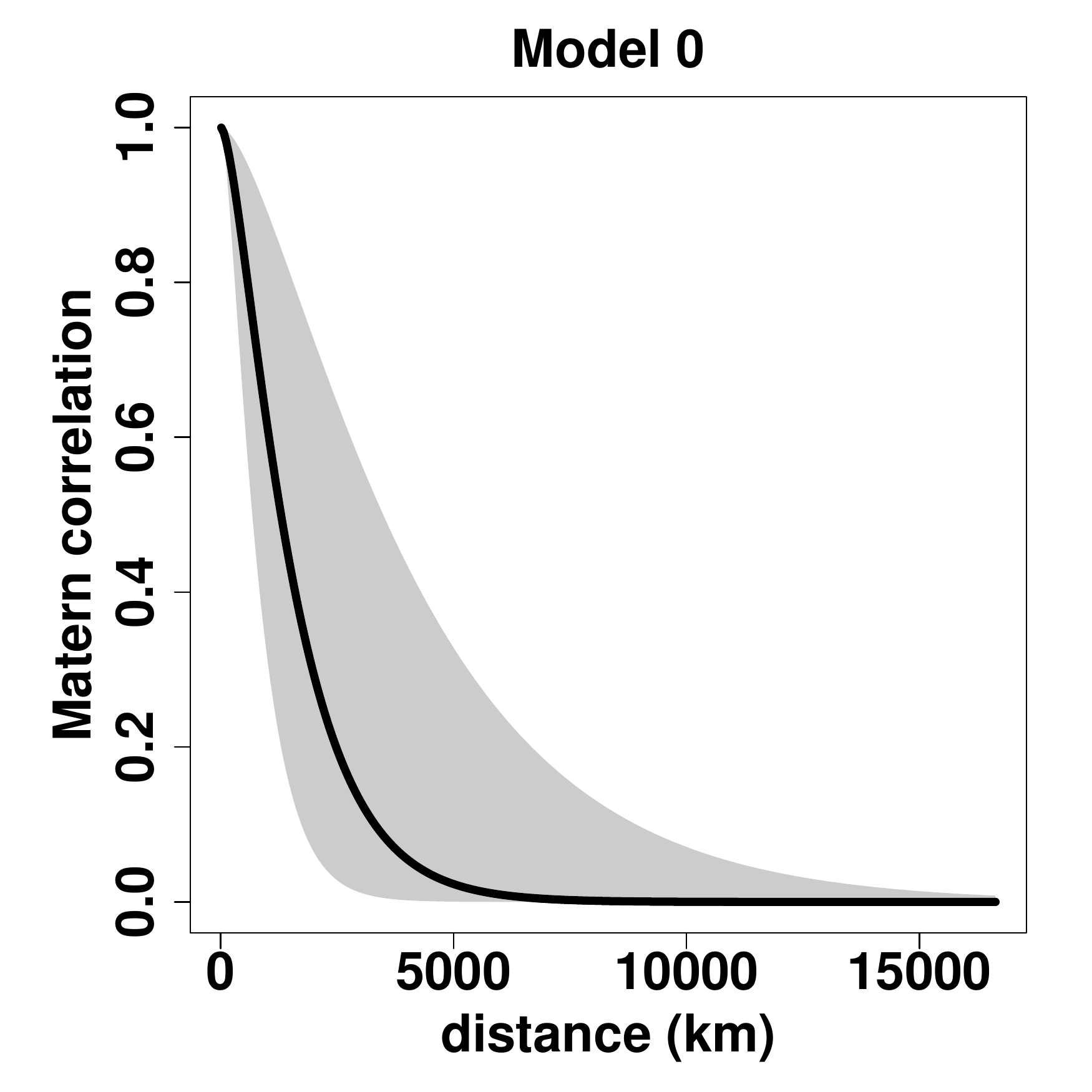}&
\includegraphics[scale=0.2]{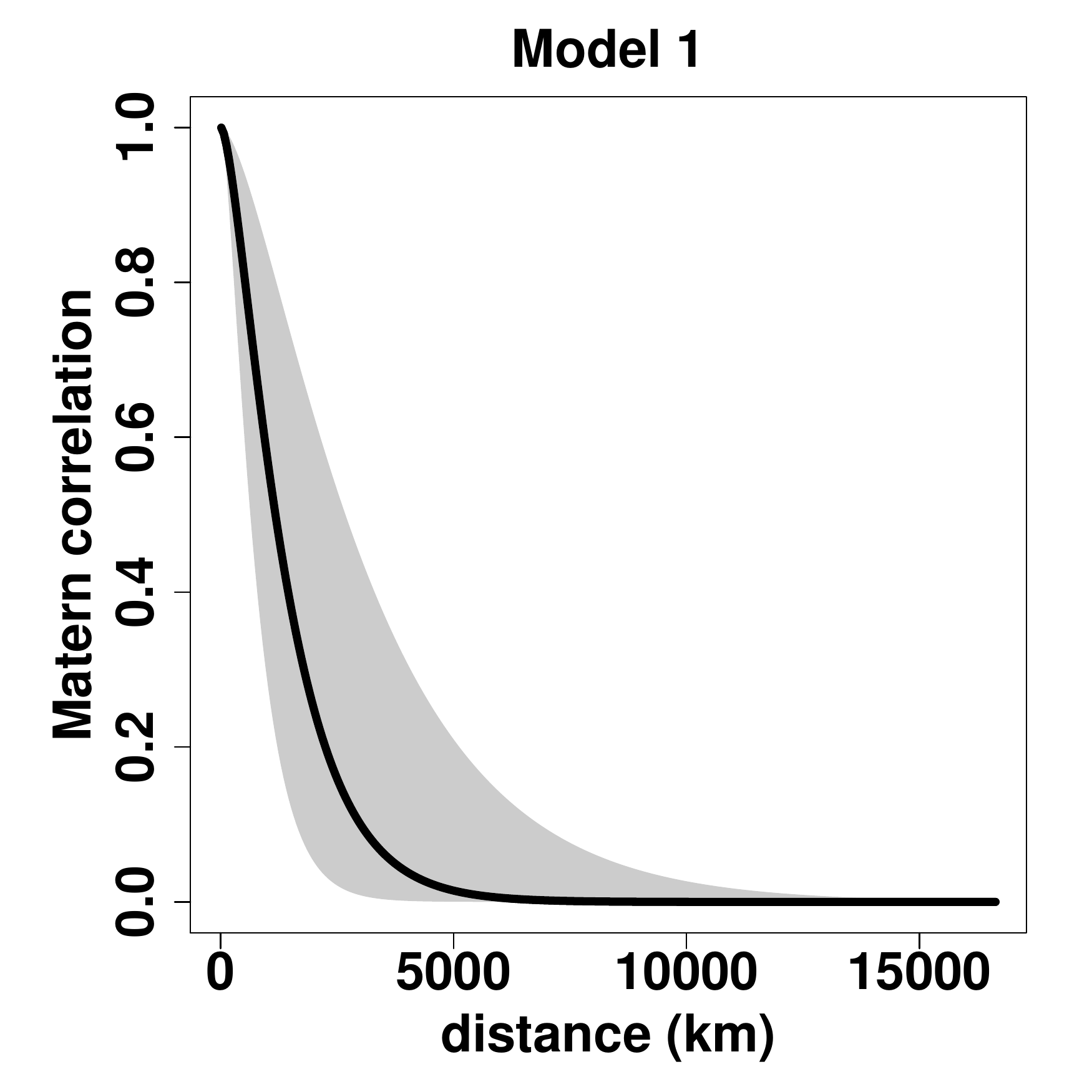}&
\includegraphics[scale=0.2]{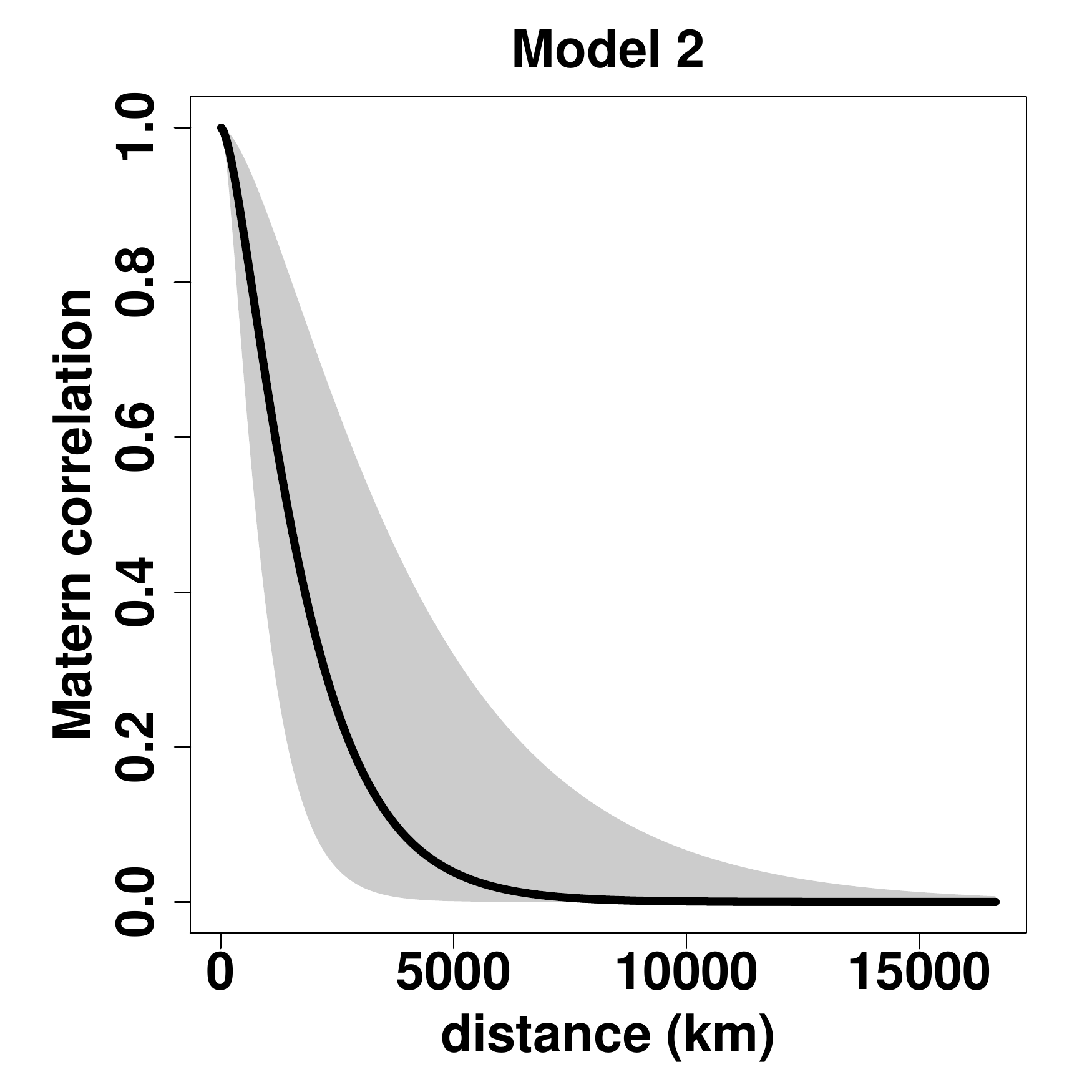}
\end{tabular}
\caption{The Mat{\'e}rn correlation function (\ref{eq:materncorr}) given the posterior estimates of $\kappa$ for Models 0, 1 and 2. In all plots, the solid lines represent (\ref{eq:materncov}) with $\kappa$ equal to its posterior median. The shaded area represents the range of (\ref{eq:materncov}) with $\kappa$ values at its posterior 2.5\% and 97.5\% quantiles. For comparison, the survey region extends roughly 14,000 km from east to west.}
\label{fig:Matern}
\end{figure}

\begin{figure}[hb]
\centering
\begin{tabular}{ccc}
\includegraphics[scale=0.2]{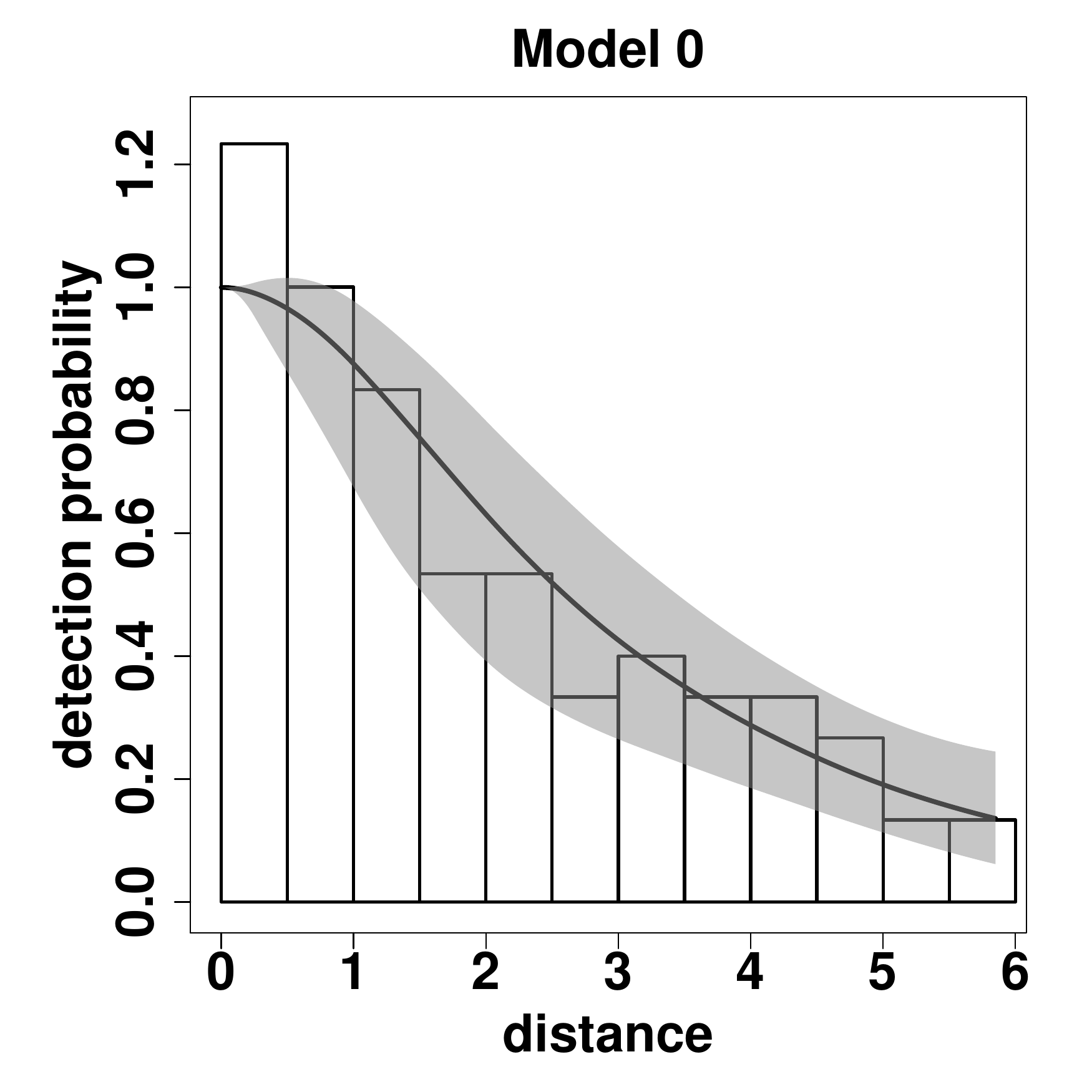}&
\includegraphics[scale=0.2]{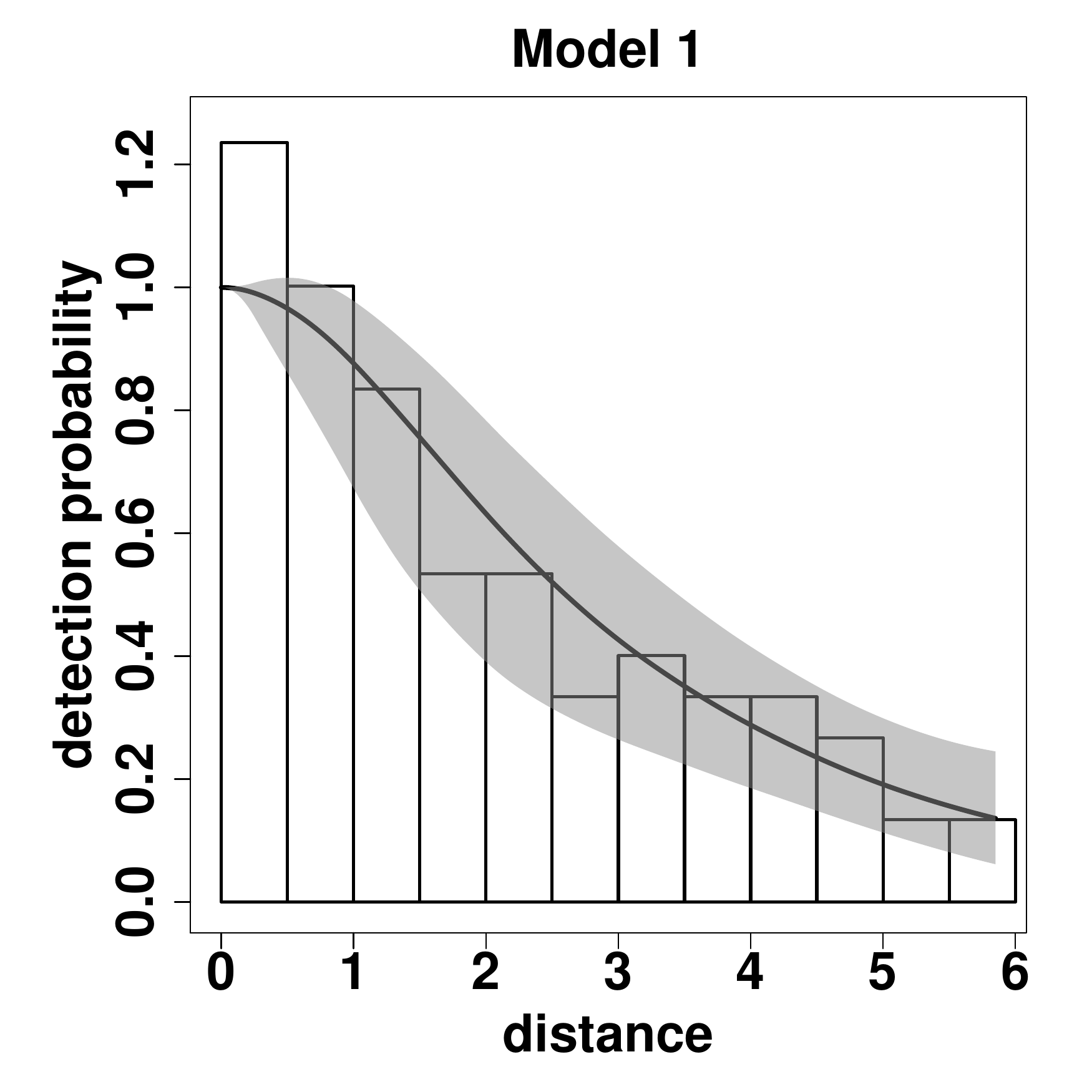}&
\includegraphics[scale=0.2]{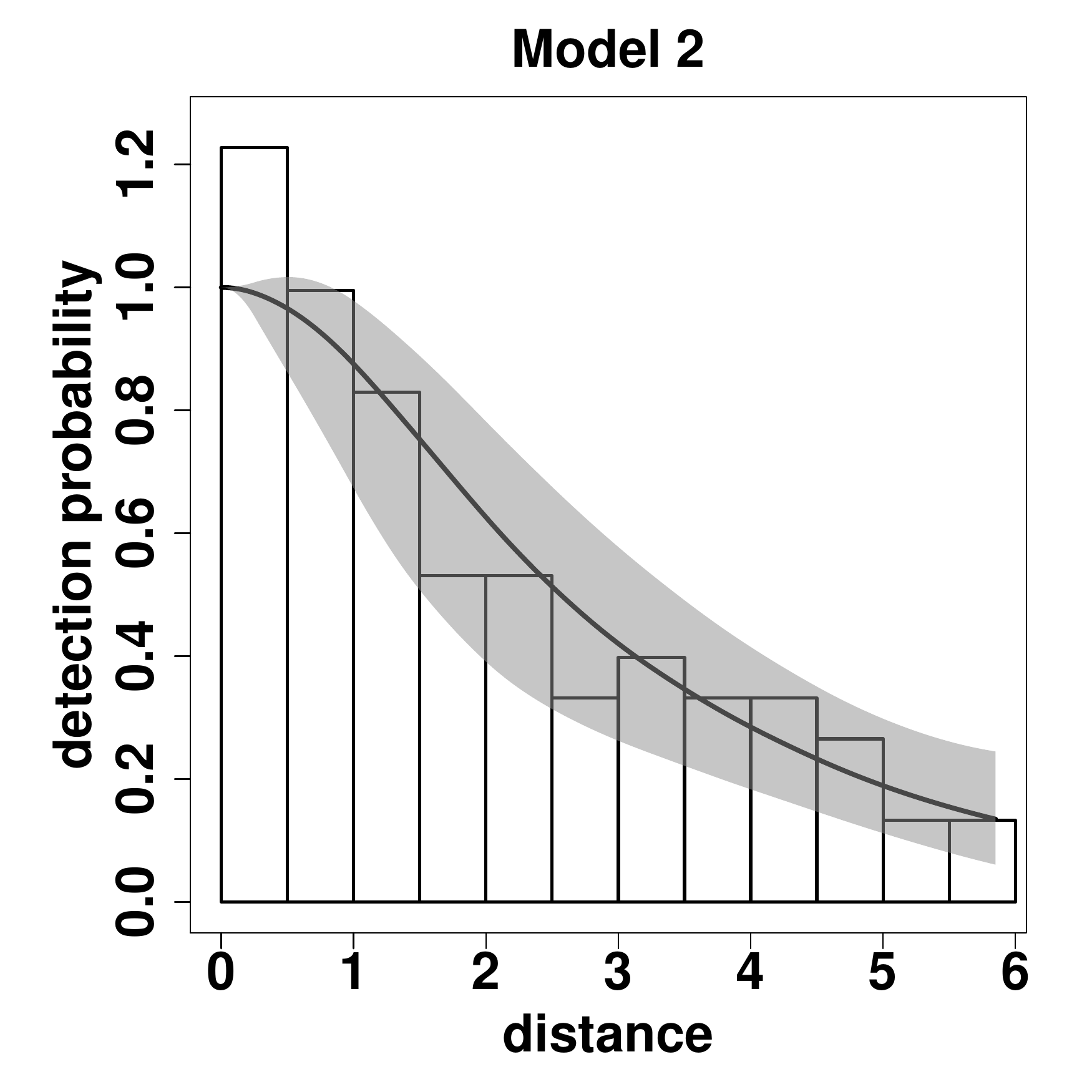}
\end{tabular}
\caption{The posterior detection function with 95\% credible band using Models 0, 1 and 2. The semi-parametric detection function is given in (\ref{eq:semiparGz}) and illustrated in Fig~\ref{fig:detectionbasis}. The posterior credible band is calculated based on the posterior distribution of $\beta$'s in (\ref{eq:semiparGz}).}
\label{fig:postgx}
\end{figure}

\clearpage
\section{Latent field and fixed effect plots}\label{appx:latenfixed}
\setcounter{figure}{0}

\begin{figure}[h!]
\centering
\includegraphics[width=.99\textwidth]{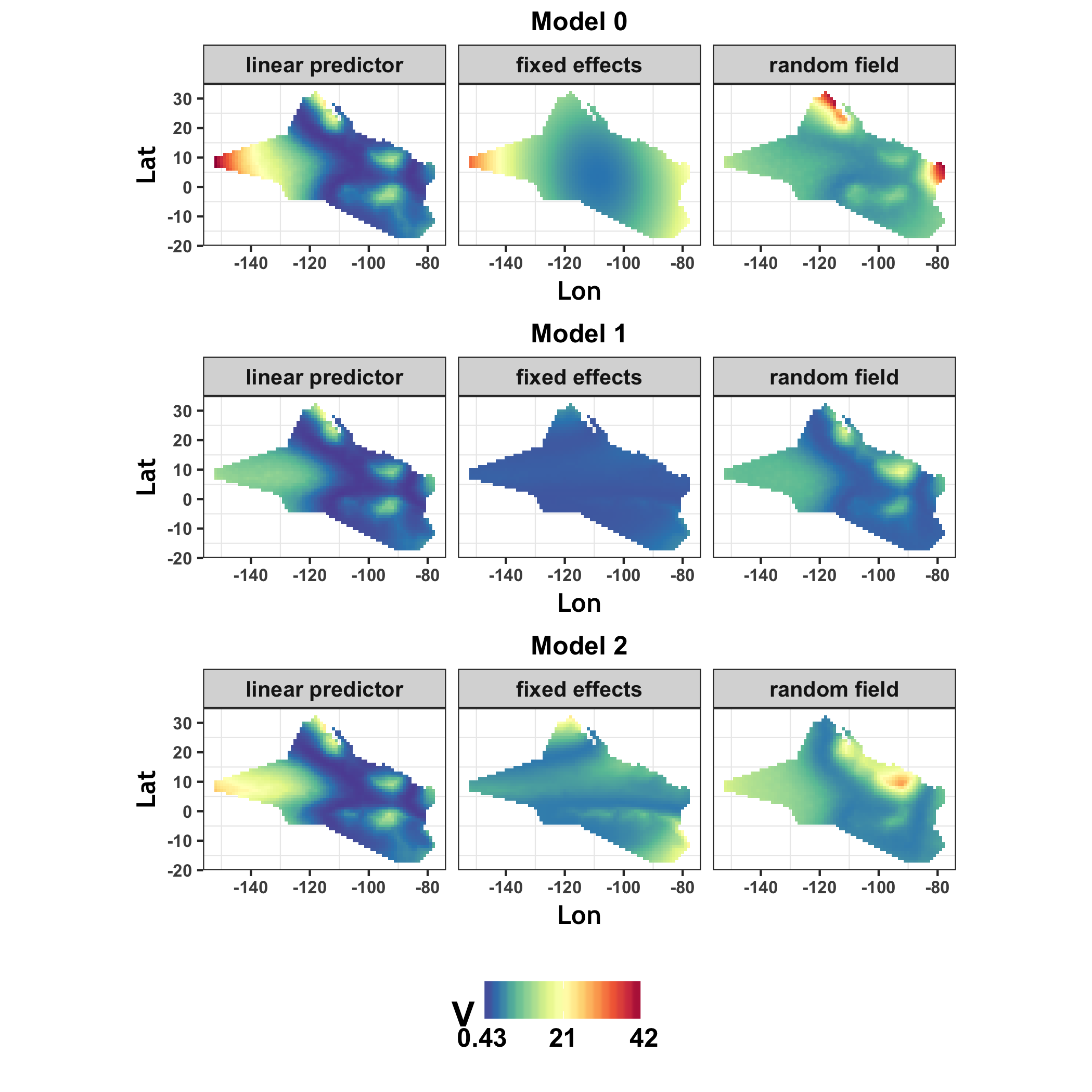}
\caption{Variability measures $V_\eta$, $V_\beta$, and $V_\xi$ for the three models. The fixed effect component of Model~0 clearly suffers from its lack of SST information.}
\label{fig:Vmodel2}
\end{figure}

\newpage

\end{document}